\documentclass[twocolumns]{emulateapj}

\shorttitle{SFHs and Sizes of Massive ETGs at z=1.2}
\shortauthors{Rettura et al.}

\begin{document}

%% LaTeX will automatically break titles if they run longer than
%% one line. However, you may use \\ to force a line break if
%% you desire.

\title{Formation epochs, star  formation histories and sizes of
  massive  early-type galaxies  in cluster  and field  environments at
  z=1.2: insights from the rest-frame UV}

%% Use \author, \affil, and the \and command to format
%% author and affiliation information.
%% Note that \email has replaced the old \authoremail command
%% from AASTeX v4.0. You can use \email to mark an email address
%% anywhere in the paper, not just in the front matter.
%% As in the title, use \\ to force line breaks.

%\author{Alessandro Rettura\altaffilmark{1}, Piero Rosati\altaffilmark{2}, Mario Nonino\altaffilmark{3}, Robert A. E. Fosbury\altaffilmark{4}, Raphael Gobat\altaffilmark{2}, Nicola Menci\altaffilmark{5}, Veronica Strazzullo\altaffilmark{6}, Ricardo Demarco\altaffilmark{1}, Holland C. Ford\altaffilmark{1}, Simona Mei\altaffilmark{7}}

\author{Alessandro Rettura\altaffilmark{1}, P. Rosati\altaffilmark{2}, M. Nonino\altaffilmark{3}, R. A. E. Fosbury\altaffilmark{4}, R. Gobat\altaffilmark{2}, N. Menci\altaffilmark{5}, V. Strazzullo\altaffilmark{6},  S. Mei\altaffilmark{7}, R. Demarco\altaffilmark{1}, H. C. Ford\altaffilmark{1}}

\altaffiltext{1}{Department of Physics and Astronomy, Johns Hopkins University, Baltimore, MD, USA}
\altaffiltext{2}{ESO - European Southern Observatory, Garching bei Muenchen, D- 85748, Germany}
\altaffiltext{3}{INAF - Osservatorio Astronomico di Trieste, via G.B. Tiepolo 11, 34131 Trieste, Italy}
\altaffiltext{4}{ST-ECF - Karl Schwarzschild straase, 2, Garching bei Muenchen, D- 85748, Germany}
\altaffiltext{5}{INAF - Osservatorio Astronomico di Roma, via di Frascati 33, I-00040, Monteporzio, Italy}
\altaffiltext{6}{National Radio Astronomy Observatory, P.O. Box O, Socorro, NM 87801.}
\altaffiltext{7}{GEPI, Observatoire de Paris, Section de Meudon, Meudon Cedex, France}

%\and

%\author{R. J. Hanisch\altaffilmark{5}}
%\affil{Space Telescope Science Institute, Baltimore, MD 21218}

%% Notice that each of these authors has alternate affiliations, which
%% are identified by the \altaffilmark after each name.  Specify alternate
%% affiliation information with \altaffiltext, with one command per each
%% affiliation.

%\altaffiltext{1}{Visiting Astronomer, Cerro Tololo Inter-American Observatory.
%CTIO is operated by AURA, Inc.\ under contract to the National Science
%Foundation.}
%\altaffiltext{2}{Society of Fellows, Harvard University.}
%\altaffiltext{3}{present address: Center for Astrophysics,
%    60 Garden Street, Cambridge, MA 02138}
%\altaffiltext{4}{Visiting Programmer, Space Telescope Science Institute}
%\altaffiltext{5}{Patron, Alonso's Bar and Grill}

%% Mark off your abstract in the ``abstract'' environment. In the manuscript
%% style, abstract will output a Received/Accepted line after the
%% title and affiliation information. No date will appear since the author
%% does not have this information. The dates will be filled in by the
%% editorial office after submission.

\begin{abstract}
  We  derive stellar  masses,  ages and  star  formation histories  of
  massive early-type  galaxies in the  z=1.237 RDCS1252.9-2927 cluster
  and compare  them with those  measured in a  similarly mass-selected
  sample  of field contemporaries  drawn from  the GOODS  South Field.
  Robust  estimates of these  parameters are  obtained by  comparing a
  large  grid of  composite stellar  population models  with  8-9 band
  photometry in the rest-frame NUV,  optical and IR, thus sampling the
  entire  relevant  domain  of   emission  of  the  different  stellar
  populations.  Additionally, we present new, deep $U$-band photometry
  of both  fields, giving  access to the  critical FUV  rest-frame, in
  order to constrain empirically  the dependence on the environment of
  the  most recent  star  formation processes.   We  also analyze  the
  morphological properties  of both samples to  examine the dependence
  of their scaling  relations on their mass and  environment.  We find
  that early-type galaxies, both in the cluster and in the field, show
  analogous  optical morphologies,  follow comparable  mass  vs.  size
  relation, have congruent average  surface stellar mass densities and
  lie  on the  same Kormendy  relation.  We  also that  a  fraction of
  early-type galaxies  in the field employ  longer timescales, $\tau$,
  to assemble  their mass than their cluster  contemporaries. Hence we
  conclude that, while the  formation epoch of early-type only depends
  on their mass, the environment does regulate the timescales of their
  star  formation  histories.   Our  deep  $U$-band  imaging  strongly
  supports  this conclusions.  I  shows that  cluster galaxies  are at
  least  0.5 mag fainter  than their  field contemporaries  of similar
  mass and optical-to-infrared colors,  implying that the last episode
  of star formation must have happened more recently in the field than
  in the cluster.

\end{abstract}

%% Keywords should appear after the \end{abstract} command. The uncommented
%% example has been keyed in ApJ style. See the instructions to authors
%% for the journal to which you are submitting your paper to determine
%% what keyword punctuation is appropriate.

\keywords{galaxies: clusters: individual: RDCS1252.9-2927 --- galaxies: high-redshift --- galaxies: fundamental parameters
--- galaxies: evolution --- galaxies: formation --- galaxies: elliptical --- cosmology: observations}

%% From the front matter, we move on to the body of the paper.
%% In the first two sections, notice the use of the natbib \citep
%% and \citet commands to identify citations.  The citations are
%% tied to the reference list via symbolic KEYs. The KEY corresponds
%% to the KEY in the \bibitem in the reference list below. We have
%% chosen the first three characters of the first author's name plus
%% the last two numeral of the year of publication as our KEY for
%% each reference.

%% Authors who wish to have the most important objects in their paper
%% linked in the electronic edition to a data center may do so by tagging
%% their objects with \objectname{} or \object{}.  Each macro takes the
%% object name as its required argument. The optional, square-bracket 
%% argument should be used in cases where the data center identification
%% differs from what is to be printed in the paper.  The text appearing 
%% in curly braces is what will appear in print in the published paper. 
%% If the object name is recognized by the data centers, it will be linked
%% in the electronic edition to the object data available at the data centers  
%%
%% Note that for sources with brackets in their names, e.g. [WEG2004] 14h-090,
%% the brackets must be escaped with backslashes when used in the first
%% square-bracket argument, for instance, \object[\[WEG2004\] 14h-090]{90}).
%%  Otherwise, LaTeX will issue an error. 

\section{Introduction}

The  description of galaxy  formation and  evolution becomes  far more
complicated if one  considers that a galaxy is  not an isolated island
universe. %In  1755  {\it  Allgemeine  Naturgesichte}  Immanuel  Kant
%correctly deduced  that the Milky  Way was a  large disk of  stars and
%suggested the  possibility that other nebulae might  also be similarly
%large  structures in mutual  interaction. Nowadays 
Many  observations have  shown  that  galaxies are  in  fact parts  of
groups, clusters  and super-clusters,  and that their  properties are
correlated with  the environment in  which they live.  Several authors
have shown  that the  local and the  large-scale environments  play an
important  role in  determining many  galaxy properties,  such  as star
formation rate, gas  content and morphology \citep{Kodama01,Balogh02}.
Several mechanisms have been proposed by theorists to account for these
effects,  such as ram  pressure stripping,  mergers and  tidal effects
\citep{Gunn72, Dressler97, Moore96,Moore98,Moore99}.

More than half of all stars in the local Universe are found in massive
spheroids (e.g., \citet{Bell03}).  To derive the assembly  history of the bulk of the stellar
mass in the Universe it is thus fundamental to understand how and when
the  early-type  galaxies  (hereafter,  ETGs)  built  up  their  mass.
Several  studies  at $1<  z  < 1.5$  indicate  that  the most  massive
galaxies in the  field ($M_{stars} > 10^{11} M\odot$)  may also be the
oldest             at             a            given             epoch
\citep{Cimatti04,Fontana04,Saracco04,Treu05,Juneau05,Tanaka05}.  Other
studies have  shown that the massive early-type  cluster galaxies have
evolved  mainly passively  since $z  \sim 1.0$  and that,  since then,
field   galaxies  have   evolved   as  slowly   as  cluster   galaxies
\citep{Bernardi98,    vandokkum96,   vandokkum01,    Treu99,   Treu01,
  Kochanek00, vandokkumstanford03,  Strazzullo06, Depropris07}.  These
studies  imply an  epoch  of massive  early-type  galaxy formation  at
redshifts well  beyond $z  = 2$.  At  earlier epochs, an  abundance of
dusty  and star forming  galaxies are  found, which  appear to  be the
progenitors of massive ETG \citep{Adelberger05}.  Contrary to what is
observed   in  low   redshift   clusters,  observations   of  $z   >2$
proto-clusters  have  shown that  the  quiescent  red sequence,  which
traces  the   passively  evolving  ellipticals,  has   yet  to  appear
\citep{Kurk04}.  This supports a paradigm where a more rapid evolution
in denser environments is occurring a $z \sim 2$.  More recent studies
\citep{Kodama07, Zirm08} reveal the  appearance of the red sequence of
galaxies by $2< z <3$, although with a large scatter.

Thus, galaxy cluster samples at $1.0 <  z < 2.0$ provide a key link to
the more  active epoch at  $z>2$ where proto-clusters  around powerful
high  redshift  radio galaxies  are  not  yet  populated by  passively
evolving ETG  \citep{Kurk04}. Cluster galaxies have  evolved and were
more luminous and bluer  at high redshift (e.g.  \citet{vandokkum98}).
Other observational results, e.g. \citet{Labbe05}, \citet{Papovich06},
\citet{Kriek06}, suggest that  by $z \sim 2$ we  are entering the star
formation  epoch  of massive  ETGs,  leading  us  to ask  whether  the
properties of  ETG at intermediate redshift are  consistent with this
interpretation.  On the basis of  these studies, the evidence seems not
only to indicate an epoch of massive early-type galaxy formation at $z
>  2$ but  also that  in the  range $1.0  < z  < 2$,  the environmental
effects start to become more substantial.

In  fact, the most  distant clusters  known to  date (all  at $z<1.5$)
provide   the   strongest   test   of  model   predictions.   Relevant
investigations include the observation of the aftermath of an off-axis
merger in  XMMXCS2215.9-1738 \citep{Hilton07}  at z=1.45, a  tight red
sequence  of  ETG  at   $0.8<  z  <  1.3$  \citep{Rosati04,  Blake03,
  Blakeslee06, Lidman04,  Mei06a, Mei06b}, a  slowly evolving $K$-band
luminosity  function  at  odds  with  hierarchical  merging  scenarios
\citep{Toft04,  Strazzullo06},   and  a  tight   and  slowly  evolving
Fundamental  Plane (hereafter,  FP) out  to z=1.25  \citep{Hol05} have
been found.  Intriguingly,  \citet{Steidel05} have found that galaxies
in a proto-cluster environment at $z=2.3$ have mean stellar masses and
inferred ages that are $\sim 2$ times larger than identically selected
galaxies outside of the structure.

A long-standing prediction of hierarchical  models is that ETG in the
field are younger for a given  mass than those in cluster cores, since
galaxy    formation    is    accelerated   in    dense    environments
\citep{Diaferio01, Delucia04}. Studies at low redshift, using chemical
abundance  indicators  \citep{Bernardi06} or  the  analysis of  fossil
record  data  via line  strength  indices \citep{Thomas05,  Clemens06,
  Sanchez06} suggest  that star formation in  low density environments
was delayed  by 1-2 Gyr.  FP studies at  $z \simeq 1$ have  shown that
massive  ETG in the field  and in  clusters  ($M_{*} >  3 \cdot  10^{10}
M_{\odot}$)  share the  same FP  evolution ($M/L$  vs.  $z$)  and have
approximately similar ages (within  $\sim 0.4$ Gyr) and star formation
histories (e.g., \citet{vandokkum07}, \citet{dSA06}).  FP studies have
also shown that $M/L$ ratios  of massive cluster and field ETG evolve
slowly and regularly and that there is evidence that low-mass galaxies
evolve       faster      than      high-mass       galaxies      (e.g,
\citet{Hol05,Jorgensen06,Treu05,vdW05,dSA05}).   This  so-called  {\it
  downsizing}  effect  is at  odds  with  earlier semi-analytic  model
predictions  \citep{Baugh96,   Kauffmann98,  Somerville04}  (see  also
\citet{Renzini06}), although  it can be reconciled in  the most recent
versions  based  on  $\Lambda  CDM$  cosmogony  \citep{Delucia06}  and
assuming that dry-mergers between  non star-forming ETG may occur and
build up  the massive early-type  galaxy population \citep{Khochfar03,
  Bell05}.

The majority of the above mentioned studies have focused on rest-frame
optical and/or infrared spectrophotometric data.  However, the optical
spectrum  remains largely unaffected  by moderate  amounts of  past or
recent  star formation.   More  recently, deep  optical surveys  (e.g.
FIRES \citep{Franx03}, GOODS \citep{Giava04}, COMBO-17 \citep{Wolf04},
MUSYC  \citep{Gawiser06}) have  provided access  to the  rest-frame UV
spectrum beyond $z\sim 0.5$, enabling more in-depth studies of the ETG
population.   Studies of  local  ETG have  revealed  the existence  of
relatively young stellar populations.  Such fossil record observations
of  absorption  line-strengths  \citep{Trager00, Thomas04}  find  that
stellar  populations younger than  $\sim 5$Gyr  (i.e. which  must have
formed between  $z \sim  1$ and the  present-day) are common  in ETGs.
Furthermore,  a  significant  fraction  of  z  $\leq  0.1$  ETG  show
relatively blue NUV-optical  colors \citep{Kaviraj05}, within extended
disks  \citep{Kauffmann07}, indicating  star-formation  over the  past
Gyr.  The inferred recent  star-formation amounts to only $\sim1$\% of
the total stellar mass.  In a more recent study, \citet{Kaviraj07} have
also shown  that a significant  fraction of $0.5  < z < 1.0$  ETG show
relatively blue NUV colors,  indicating star-formation over the past 1
Gyr.  At  slightly higher redshift,  spectroscopic studies at  $z \sim
1.2$  have  shown  that  both  the brightest  ETG  of  RDCS1252.9-2927
\citep{Demarco07}   and   the  massive   ETG   in   the  GDDS   fields
\citep{Leborgne06}  show  evidence  for  recent  (i.e.,  within  1Gyr)
star-formation on  the basis  of prominent post-starburst  features in
galaxy spectra (e.g., $H_{\delta}$ absorption line).

\begin{figure*}
\epsscale{1.}
\plotone{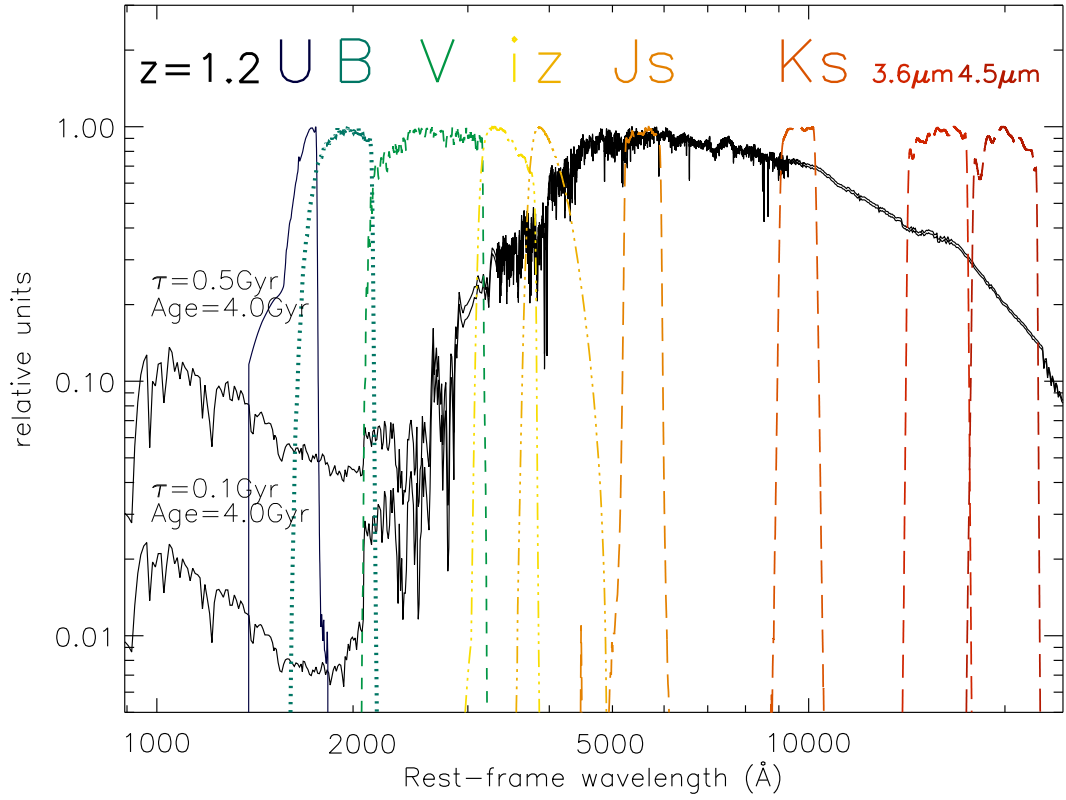}
\caption{BC03  composite  stellar  population  models  of  4.0Gyr  old
  models,  of different  $\tau$s. The  colored lines  are  the filter
  transmission curves  of the observing  bands we use  throughout this
  work, shifted to our sample rest-frame.}
\label{sed_model}
\end{figure*}

However, since little is known about the dependence on the environment
of  the recent  star-formation rates  (hereafter, SFR)  in ETG  at $z
\simeq  1.2$, in  this  work,  we complement  our  modeling of  galaxy
Spectral  Energy Distributions  (hereafter,  SEDs), from  NUV to  NIR,
presenting result  of deep observations  of ETG in cluster  and field
obtained with  {\it VLT}/VIMOS in  the $U$-band filter,  that directly
probes the  FUV regime  at the  redshift of our  samples, as  shown in
Fig~\ref{sed_model}.

By studying stellar population ages  at $z=1.2$, we provide a key test
of the  paradigm of  an accelerated evolution  in the  highest density
environments. For galaxies observed at $z \leq 1$ most of their difference could have been
smoothed  out  by  billion  years  of mostly  passive  evolution.   By
comparing stellar masses, ages  and inferred star formation histories
of cluster  ellipticals with  their field contemporaries,  we directly
test  the  prediction  that  the cluster  environment  should  display
accelerated evolution, resulting in larger masses and ages.

The primary observational goal of this  work is to use {\it HST}/ACS
in     the     rest-frame     Near-UV     (hereafter,     NUV)     and
optical\footnote{wavelength  ranges  which  are  known to  be  stellar
  population age sensitive},and  {\it VLT}/ISAAC, {\it Spitzer}/IRAC in
the  rest-frame near-IR  (hereafter, NIR)\footnote{a  wavelength range
  which is known to be strongly correlated with the underlying stellar
  mass  \citep{Gavazzi96}}  to  measure  stellar population  ages  and
masses   for    ETG   in   the    z=1.237   RDCS1252.9-2927   cluster
\citep{Rosati04}  and  compare them  to  those  measured on  similarly
selected  sample of field  contemporaries drawn  from the  GOODS South
Field.   This allows  us  to directly  analyze  the  entire
relevant spectral energy distribution  of the different stellar populations,
enabling us to improve constraints on galaxy ages, masses and star formation
histories in both environments at $z \simeq 1.2$.

We note that in an accompanying paper, \citet{Gobat08}, we also compare
the  coadded spectroscopic  information available  on both  samples of
ETG with a large grid of composite stellar population models.

We also analyze  the morphological properties of ETG  in the field and
in  the  cluster.  By  studying  scaling  relations  in this  relevant
redshift range we can trace back where the majority of stars formed as
a function of the
environment and stellar mass.\\

The structure  of this  paper is as  follows.  The description  of our
data-sets, cataloging and sample selection is described in \S 2. In \S 3
we describe our methods in  deriving galaxy sizes and morphologies as
well as  inferring ages and  masses from stellar  population analysis.
The  results  of  our  study are  discussed in  \S  4,  while  in  \S
\hspace{0.05cm} 5 we summarize our conclusions.

We assume a $\Omega_{\Lambda} = 0.73$, $\Omega_{m} = 0.27$ and $H_{0}
= 71\ \rm{km} \cdot \rm{s}^{-1} \cdot \rm{Mpc}^{-1}$ flat universe
\citep{Spergel03}, and use magnitudes in the AB system throughout this
work.

\section{Description of the data}

This work is based on spectroscopic and photometric data of two fields
which  have  extensive spectral  coverage  over the  wavelength
range  $0.4-8\mu$m: the  Chandra  Deep Field  South (hereafter,  CDFS)
observed  by  the  Great  Observatories  Origin  Deep  Survey  (GOODS,
\citep{Giava04})  and  the field  around  the  X-ray luminous  cluster
RDCS1252.9-2927 at $z=1.237$ \citep[hereafter  CL1252;][]{Rosati04}.

The  archival data  for the  CDFS comprise  deep imaging  in  8 bands,
\textit{HST}/ACS ($B_{F435W}$,  $V_{F606W}$, $i_{775W}$, $z_{F850LP}$)
\citep{Giava04}, \textit{VLT}/ISAAC, ($J$,$K_s$)  (Retzlaff et al., in
prep.),  \textit{Spitzer}/IRAC ($3.6 \mu\mbox{}m$,  $4.5 \mu\mbox{}m$)
\footnote{CDFS imaging  and spectroscopic data  are publicly available
  through        the        GOODS       collaboration        web-site:
  http://www.stsci.edu/science/goods/.}, as well as spectroscopic data
taken  with  \textit{VLT}/FORS2 300I  grism  by  the ESO-GOODS  survey
\citep{Vanzella05,Vanzella06} and the K20 survey \citep{Cimatti02}.

The  archival  data for  CL1252  consist of  deep  imaging  in 9  bands,
\textit{VLT}/FORS2      ($B$,     $V$,      $R$),     \textit{HST}/ACS
($i_{F775W}$,$z_{F850lp}$)     \citep{Blake03},     \textit{VLT}/ISAAC
($J_s$,$K_s$)     \citep{Lidman04},     \textit{Spitzer}/IRAC    ($3.6
\mu\mbox{}m$, $4.5 \mu\mbox{}m$), as  well as spectroscopic data taken
with \textit{VLT}/FORS2 and already published in \citet{Demarco07}.

\subsection{Cataloging of observations and sample selection}

This work builds on the analysis performed on the same data-sets by
\citet{Rettura06},  where photometric-stellar (hereafter,  simply {\it
  stellar}) masses and  dynamical masses of $z\sim 1$  ETG with known
velocity  dispersion  measurements were  analyzed  for  both the  CDFS
and CL1252.   We  refer  to  the above  mentioned  paper for  more
details  about  data  reduction  and cataloging  (e.g.,  photometrical
errors, PSF-matched photometry).

The resulting data-sets for the cluster and the field have homogeneous
depths and  wavelength coverage,  allowing the application  of similar
selection criteria for both samples. The data allow the reconstruction
of galaxy SED  by sampling the entire relevant  spectrum range emitted
by   all  the  different   stellar  populations.    As  is   shown  in
Fig~\ref{sed_model}, our  SED-fitting analysis  is based on  data from
the  NUV rest-frame  ($B$-band observations  are centered  at $\lambda
\sim  2000$\AA\mbox{} at  z$\simeq  1.2$) through  the NIR  (IRAC/$4.5
\mu$m imaging probes  the $\sim 2\mu m$ rest-frame  at the redshift of
CL1252).

The availability of 8 to 9 passbands spanning such a large wavelength
range enables the estimate  of  accurate  stellar  masses of  ETGs
\citep{Rettura06}  and makes it possible to  compare  stellar population
properties  of   homogeneously  selected  samples  of   ETG  in  both
environments.  In addition,  high  quality  \textit{HST}/ACS
$z_{F850LP}$-band imaging  enables the study of  their morphologies in
great detail.

Throughout this work, we  compare morphological and stellar population
properties of  cluster galaxies with those  shown by a  sample of {\it
  field contemporaries}  drawn from  the spectroscopic surveys  in the
redshift  range $z=  1.237 \pm  0.15$.   Although photometric-redshift
selected  samples are  widely employed  in the  literature, we  do not
favor this approach,  as we believe it may result  in the pollution of
the  samples by  a large  fraction of  redshift outliers,  which could
adversily affect our conclusions.

The   depth  of   the  {\it   VLT}/ISAAC  images   and   the  extended
multi-wavelength  data for both  fields allows  us to  define complete
mass-selected samples.  In  an accompanying paper, \citet{Gobat08}, we
study the  relative photometric and spectroscopic  completeness of our
CDFS and CL1252 mass-selected samples.  The reader is referred to that
paper for  more details.  Here, we note  that photometric completeness
is obtained  if we  limit our analysis  to stellar masses  larger than
$M_{lim}= 5 \cdot 10^{10} M_{\odot}$.

A  selection   of  CL1252  ETG  along  the   cluster  red-sequence  is
efficiently provided  by a color selection of  $i_{775}-z_{850} > 0.8$
\citep{Blake03}.   In the  spectroscopic  sample of  \citet{Demarco07}
there  are 22  red sequence  galaxies ($i_{775}-z_{850}  >  0.8$) with
$M_{*} > M_{lim}$, of which 18 are classified as passive ETGs, with no
emission line  in their observed spectra.  For  the corresponding CDFS
field  sample, the  same  criteria yield  27  ETG in  CDFS with  FORS2
spectra giving redshift in the range $z= 1.237 \pm 0.15$.  From visual
morphological  analysis,   following  the  classification   scheme  of
\citet{Blake03},  the vast  majority  of the  selected  ETG also  have
typical elliptical or lenticular morphologies

Comparing, as  a function of  stellar mass, each  spectroscopic sample
with  its  corresponding photometric-redshift  sample  (see Fig.1  and
Table 1  of \citet{Gobat08}) we find that  the spectroscopic follow-up
for CL1252 is  more complete at the low mass end  (reaching a $\sim 60
\%$ completeness by $M_{*}=M_{lim}$) than  in CDFS (reaching a $60 \%$
completeness only  by $M_{*}\simeq 2 \cdot  10^{11} M_{\odot}$).  Thus
our sample of ETG  in CDFS is likely to be more  incomplete at the low
mass  end than  the CL1252  one.  We  will return  to this  point when
discussing our results in \S 4.2.

%The depth  of the {\it VLT}/ISAAC  images of both fields  allows us to
%define complete mass-selected samples.  The CDFS and CL1252 $K_{s}$-band
%images  are photometrically  complete down  to  $K_{s} =  24$.  At  $z
%\simeq 1.2$ the $K_{s}$-band photometry  is considered a good proxy of
%the  stellar mass,  with $10^{10}  M_{\odot}$ corresponding  to $K_{s}
%\sim  23$  \citep{Strazzullo06}, which  we  take  as  a reliable  mass
%completeness limit.   On the  other hand, the  spectroscopic follow-up
%work is  limited to approximately $K_{s}  \simeq 22$ for  ETG in both
%samples, corresponding to $R_{Vega}  \sim 25$ and stellar mass, $M_{*}
%\simeq 3 \cdot 10^{10} M_{\odot}$.  Therefore we limit our analysis to
%photometric  stellar  masses larger  than  $M_{lim}=  5 \cdot  10^{10}
%M_{\odot}$ .

\subsection{VIMOS  $U$-band  photometry   of  CDFS  and  RDCS1252.9-2927
  fields.}

The new observations with  \textit{VLT}/VIMOS in the $U$-band allow us
to directly study the FUV rest-frame emission, which is very sensitive
to  recent   star  formation.   This   allows  us  to   constrain  the
instantaneous and  recent star-formation in massive ETG  as a function
of their environment.

The reader  is referred to  \citet{Nonino08} for more details  on data
reduction and cataloging. These  surveys provide deep $U$-band imaging
in the CDFS  for a total integration time of $\sim  $15h, to AB depths
of $U$= 28.27 mag (3$\sigma$, 1'' radius aperture), and in the cluster
region  for a total  integration time  of $\sim$2.5h,  to AB  depth of
$U$=27.3 mag. We note that the $\sim$1 mag difference in depth between
the  two data-sets  is not  only due  to the  difference in  the total
allocated  time, but  also to  the larger  galactic extinction  at the
location of the  CL1252 cluster ($A(V)=0.247$, compared to  the one at
the  CDFS location, $A(V)=0.026$).   Assuming an  extinction following
the \citet{Cardelli88}  relation, we estimate  a dimming of  $\sim 0.4$
mag in the VIMOS $U$-band for CL1252 with respect to CDFS.

The VIMOS U-band filter has a colour term with respect to
standard Johnson U-band filter.  However this term has been set to
zero both in CL1252 and CDFS fields in the process of catalog
creation, thus placing the reported magnitudes in the VIMOS-U system.

\section{Data Analysis}

\subsection{Derivation of Galaxy Sizes and Morphologies}

To  find structural differences  shown by  ETG  of the
same mass in different environments  we have to study their morphology
in a more quantitative way than  a simple visual analysis.  One way is
to model and compare their galaxy light distributions, which are known
to correlate with galaxy type and dynamical state. We also measure and
compare  galaxy  sizes  as  a   function  of  their  mass,  to  obtain
information on the  physical scale of the potential  well in which the
stellar mass is assembled.

We   have  used   GIM2D,   a  fitting   algorithm  for   parameterized
two-dimensional  modeling  of  surface brightness  (SB)  distribution
\citep{Simard98,Marleau98}, to  fit each galaxy  light distribution by
adopting a simple \citet{Sersic68} profile of the form:
\begin{equation}  \label{eq:sersic}
I(r)=I_{e_{n}}\cdot10^{-b_n[(r/R_{e,n})^{1/n}-1]},
\end{equation}
where  $b_n  = 1.99  n  -0.33$  \citep{Cap89},  and $R_{e,n}$  is  the
effective  radius (i.e., the  projected radius  enclosing half  of the
light). The  classical de Vaucouleurs profile  thus simply corresponds
to a S\'ersic index $n=4$  and $b_n=7.67$ in Eq.~\ref{eq:sersic}.  In
this work, we allow $n$ to span the range between 0 and 5.

GIM2D performs a  profile fit by deconvolving the  data with the point
spread function.  We model  PSFs with analytic functions from visually
selected stars in the surrounding  ($30'' \times 30''$) region of each
galaxy. We model  a different PSF for each region  in order to account
of  PSF  variations with  the  position in  the  field.   A 2D  radial
multi-gaussian  function has  been  fitted simultaneously  to tens  of
selected stars  around the galaxies  of our sample with  outputs being
stacked together  to create  a single PSF  image for each  region. The
reader is referred to \citet{Rettura06} for more details on our method
for  modeling   galaxy  PSFs  and  SB  profiles   from  {\it  HST}/ACS
images. Here we  also note  that using PSF  constructed with distortion-corrected Tiny Tim  \citep{Krist95} models results  in negligible differences (see, e.g., \citet{vdW05}, \citet{Treu05}).

The result of the bidimensional  fit is the semi-major axis $a_{e}$ of
the projected elliptical isophote  containing half of the total light,
the axis ratio $b/a$ and the S\'ersic index $n$, which we have left as
free parameters.   The effective  radius is computed  from $R_e  = a_e
\sqrt{b/a}$.   The  average surface  brightness  within the  effective
radius (in mag/arcsec$^2$) is  obtained from the absolute magnitude M:
\begin{equation}  \label{eq:mu_e}
\langle\mu\rangle_e  =   M  +  5logR_e  +  38.567,
\end{equation}
with  $R_e$  in kiloparsec.   In  order  to  obtain the  morphological
parameters  in  the  rest-frame  $B$-band,  which  is  customary  in
morphological  studies, we have  used the  {\it HST}/ACS  images taken
with the F850LP filter, since these  are very close to the $B$-band at
the  redshift of  our  galaxies for  both  the cluster  and the  field
samples.

\subsection{Derivation of stellar masses and star-formation weighted ages}

%% The displaymath environment will produce the same sort of equation as
%% the equation environment, except that the equation will not be numbered
%% by LaTeX.

Adopting a  similar approach  to \citet{Rettura06}, we  derive stellar
masses and  ages for each  galaxy in our samples  using multi-wavelength
PSF-matched aperture photometry  from 8 and 9 passbands  for the CDFS and
 CL1252 fields respectively, from  observed $B$-band to observed $4.5 \mu
m$.   For each  galaxy, we  compare  the observed  SED with  a set  of
composite stellar populations (hereafter, CSP) templates computed with
models  built with  \citet{BC03} models,  assuming  solar metallicity,
\citet{Salpeter55} Initial Mass Function (hereafter, IMF) and dust-free
models.  In  \citet{Rettura06} we did  investigate the effect  of dust
extinction on the  best-fit stellar masses by including  a fourth free
parameter,  $0.0<  E(B-V)<   0.4$,  following  the  \citet{Cardelli89}
prescription.  By  performing the  fit on  28 ETG at  $z \sim  1$, we
found that  in $ \sim 40\%$  of the cases $E(B-V)=0$  gives the best
fit.  In the remaining cases, masses  which are lower by $0.2 \pm 0.1$
dex  are  found,  with  corresponding  $E(B-V) \leq  0.2$.   This  test
supports the validity of the dust-free model assumption.

For  our CSP  models, we assume  the following  grid of
exponentially-declining star formation  history (SFH) scenarios, $\Psi
(T-t^{'}, \tau)$:

\begin{equation} \label{eq:sfhs}
\Psi (T-t^{'}, \tau) =  e^{-\frac{T-t^{'}}{\tau}} \cdot \frac{M_{\odot}}{yr},
\end{equation}

where  $0.05 \leq  \tau \leq  5$  Gyr,  $T$  is  the cosmic  time  and
$(T-t^{'})$ is  the age\footnote{The  range of acceptable
  ages for a given galaxy has  been limited by the age of the universe
  at its observed redshift.} of the stellar population model formed at
time $t^{'}$ at a SFR, $\Psi(t^{'})$.

In determining galaxy model  ages, masses and star formation histories
from SED  fitting, is important  to understand how much  our estimates
could    possibly    be    affected    by   dust    extinction,    and
``age-metallicity''\footnote{We  emply  the  working
  assumption that the most-massive  ETG have all solar metallicities.}
and ``age-SFH'' degeneracies.

\begin{figure*}
\epsscale{1.0}
\includegraphics[scale=.7]{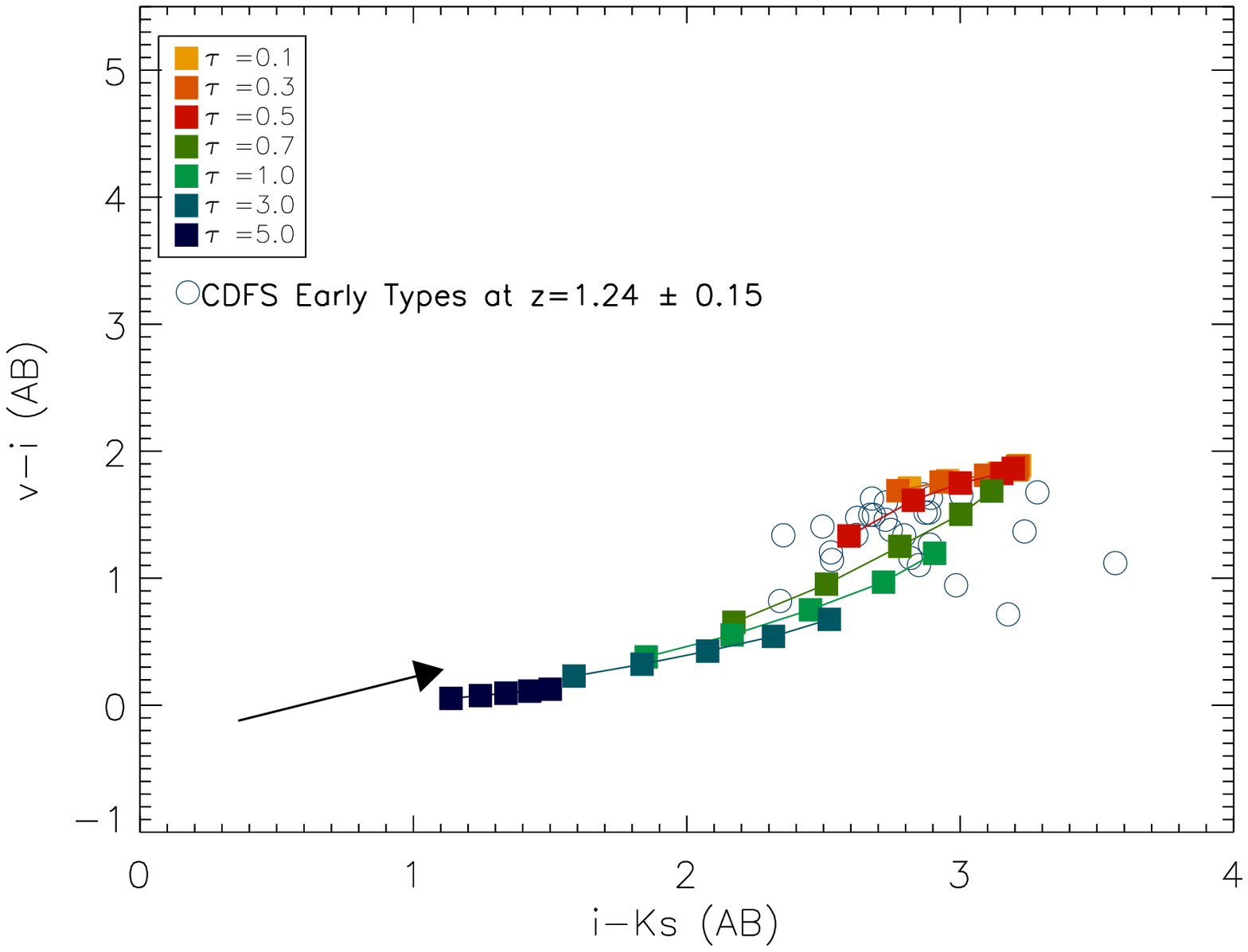}
\includegraphics[scale=.7]{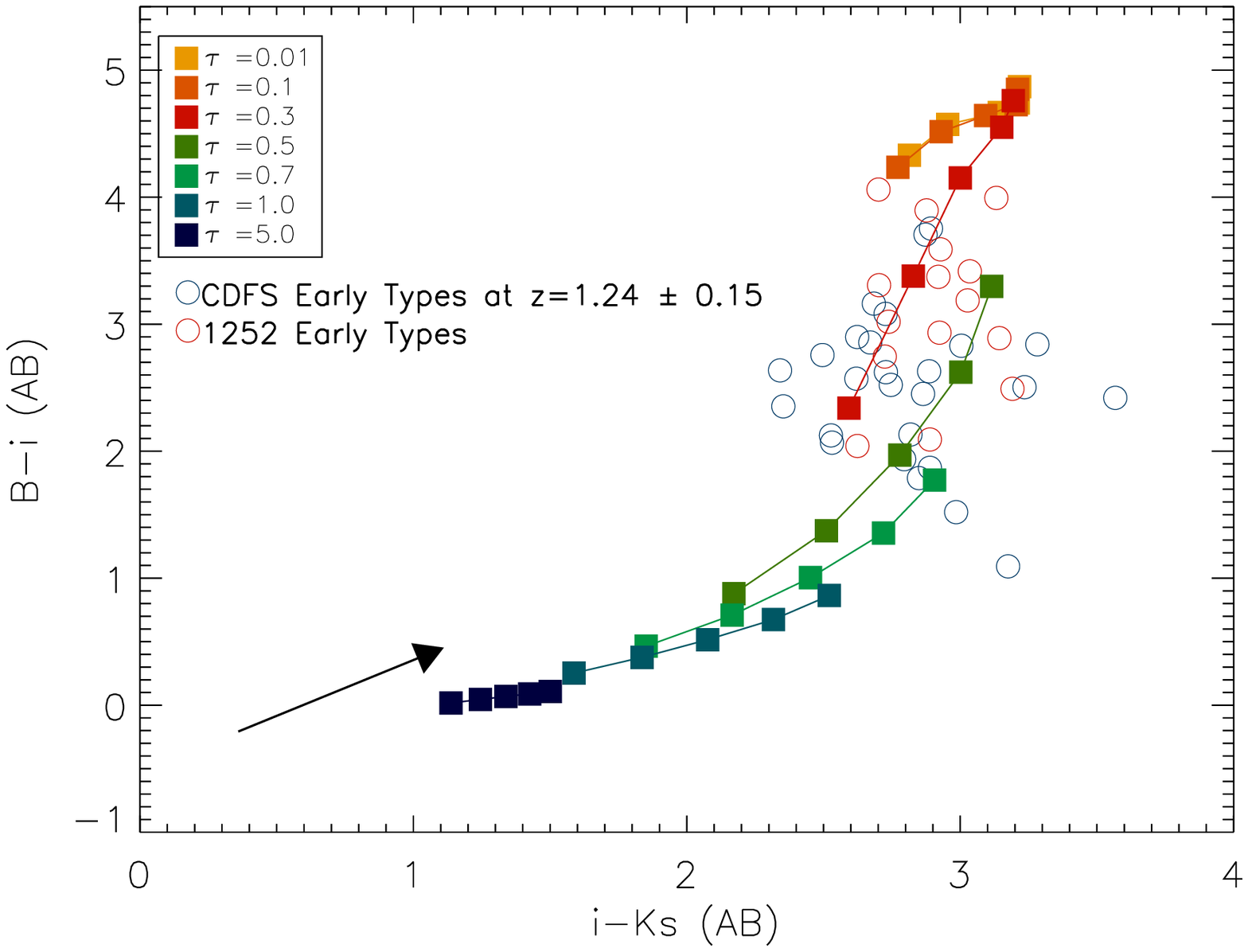}
\caption{{\it Top panel}: $i-K_s$  vs $V-i$ color-color plot of $BC03$
  composite stellar population models  at $z=1.24$ superimposed on our
  GOODS  ETG photometry  at $z=1.24  \pm  0.15$ in  blue circles.  The
  squares account for SFHs with  various $\tau$ models.  The grids are
  drawn for seven different $\tau$s  and five ages  ($2, 2.5, 3,  3.5, 4$
  Gyrs):  the  colored  lines  represent  iso-metallicity  colors  of
  $Z_{\odot}$.   The   black  arrow  indicate   an  extinction  of
  $E(B-V)=0.2$   as  parameterized   with  the   reddening   curve  of
  \citep{Cardelli89}.  {\it  Bottom left panel:}  $i-K_{s}$ vs.  $B-i$
  color-color diagram of  the same models and data: $B$-band ($\lambda_{rest} \sim 2000 \AA$)
  is mandatory to break the age-SFH degeneracy at z=1.24}
\label{colmod}
\end{figure*}

Galaxies could appear redder as a result of any of, a shorter $\tau$ ,
a larger  extinction, or  an older age,  would all transform  a galaxy
spectrum into a redder one.   This effect is simply illustrated in the
top  panel  of Fig.~\ref{colmod}.   We  show  the $i_{775}-K_{s}$  vs.
$v_{606}-i_{775}$  color-color  plot  of  BC03 CSP  models  at  z=1.24
superimposed on our CDFS ETG sample observed colors.  The squares show
the various $\tau$ models predictions.   The grids are drawn for seven
different $\tau$  and five model  ages (2, 2.5,  3, 3.5, 4  Gyrs): the
colored lines  represent iso-metallicity colors  of solar metallicity,
$Z_{\odot}$.  It is evident from this figure that it might become very
hard  to  distinguish  different  model parameters  with  SED  fitting
studies  based  on rest-frame  optical  and  infrared photometry  only
(i.e.,  $\lambda_{rest}  >$  2700   \AA).   However,  as  we  show  in
Fig.~\ref{sed_model},   the  use  of   information  coming   from  the
rest-frame UV  is crucial to  distinguish the different  parameters of
the  stellar  population modeling  (e.g.   ages  and  $\tau$).  As  we
demonstrate  in the  bottom panel  of Fig.~\ref{colmod},  by including
available  UV   rest-frame  photometry  in  the   SED  fits  ($B$-band
observed-frame\footnote{corresponding  to $\sim$  2000 \AA  \mbox{} at
  $z=1.24$}) we  are able to  break the ``age-SFH  degeneracy''.  Note
that  the  rest-frame  UV  remains  also  largely  unaffected  by  the
``age-metallicity'' degeneracy, which plagues optical studies
\citep{Worthey94}.   Hence the extensive  panchromatic method we use
maintains its age-sensitivity across a large range of masses, ages and
$\tau$, providing more reliable estimates of these parameters compared
to those obtained using optical-to-infrared studies.

To account for the  average age of the bulk of the  stars in a galaxy,
we refer throughout this  paper to {\it star-formation weighted} ages,
$\overline t$, defined as:

\begin{equation} \label{eq:age2}
\overline t(T-t^{'}, \tau) \equiv \frac{\int_{0}^{T} (T-t^{'}) \Psi(T-t^{'}, \tau) dt^{'}}{\int_{0}^{T} \Psi(T-t^{'}, \tau) dt^{'}} .
\end{equation}
Assuming $\Psi(T-t^{'}, \tau)$ as in Eq.~\ref{eq:sfhs} we obtain,
\begin{equation} \label{eq:age}
\overline t = \tau \cdot e^{-\frac{T-t^{'}}{\tau}} + (T-t^{'}) + \tau.
\end{equation}

By  comparing  each  observed  SED  with these  atlases  of  synthetic
spectra, we construct  a 3D $\chi^{2}$ space spanning  a wide range of
star formation  histories, model ages  and masses. The galaxy  mass in
stars $M_{*}$, the $\tau$ and the inferred $\overline t$ of the models
giving the lowest $\chi^2$ are  taken as the best-fit estimates of the
galaxy stellar mass,  age, and SFH timescale.  
%The  errors on the ages
%and  the masses  are estimated  as in  \citet{Rettura06}, to  which we
%refer the  reader for more details.   Here we remind  that 
We note that this
procedure results in typical errors  for galaxy ages of $\sim 0.5$Gyr,
and for $\tau$  of $\sim  0.2Gyr$. Typical  uncertainties on  the mass
determination are about $\sim 40\%$ (i.e., $0.15$ dex)  \citep{Rettura06}.

The reliability  of spectrum  synthesis models at  $\lambda_{obs} \sim
2\mu  m$  has long  been  debated  (\citet{Maraston98} and  references
therein).  In the rest-frame NIR regime, in early stages of the galaxy
evolution,  a short-duration  thermally pulsating  (TP-) AGB  phase is
known  to  be  relevant.   In  \citet{Rettura06} we  have  shown  that
PEGASE.2  \citep{Fioc97}, BC03,  and Maraston  models \citep[hereafter
M05;][]{Maraston05}  yield consistent  stellar masses  (within typical
errors of $40 \%$) for z$\sim  1$ ETGs. Therefore we do not expect our
stellar mass estimates to much depend on the actual stellar population
synthesis model adopted.  On the other  hand, in \S 4.2 we discuss the
effect on the galaxy ages of the use of other models such as M05.

\begin{figure*}
\epsscale{1.}
\plotone{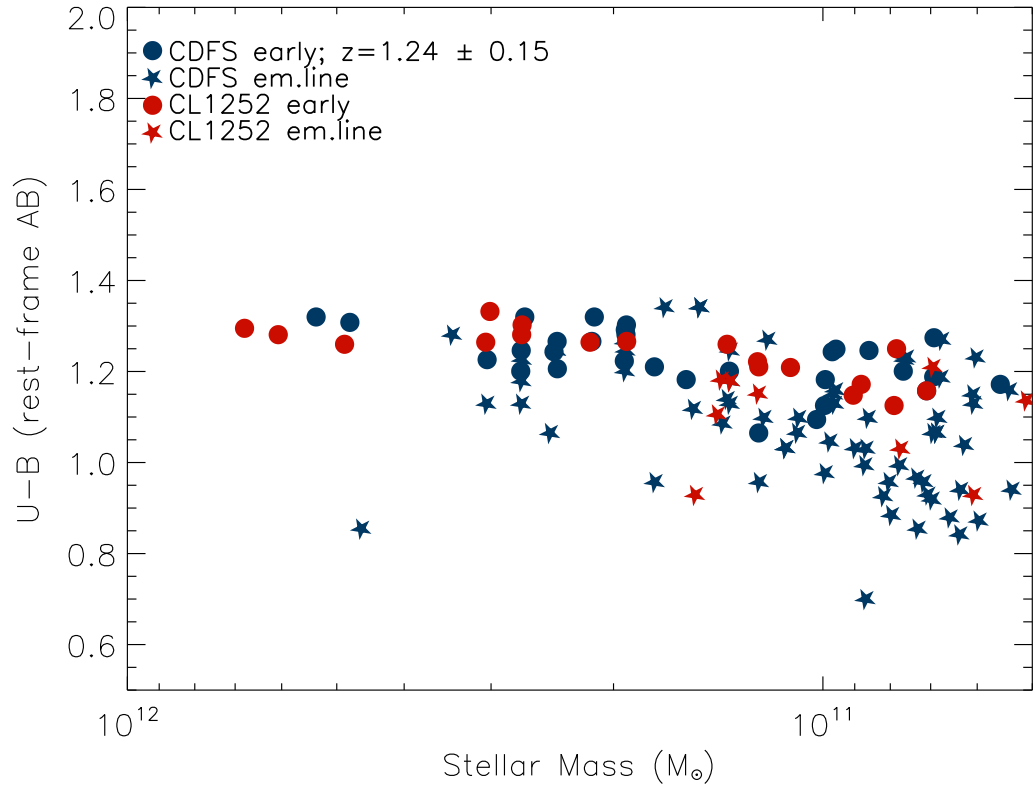}
\caption{$U-B$ color  - mass Diagram  of the mass-selected  samples of
  CL1252  early-  (filled  red  circles)  and  late-type  (red  stars)
  galaxies as well  as of CDFS field early-  (filled blue circles) and
  late-type (blue stars) galaxies.   Uncertainties in the stellar mass
  are $\sim 0.15$  dex. Field ETG galaxies are  distributed around the
  cluster red-sequence, although with a larger scatter}
\label{UMB_mass}
\end{figure*}

In  Fig.~\ref{UMB_mass}   we  plot  the  stellar   mass  versus  $U-B$
rest-frame  color  diagram  of  the mass-selected  samples  of  CL1252
cluster early- (filled red circles) and late-type (red stars) galaxies
as well  as of CDFS field  early- (filled blue  circles) and late-type
(blue stars)  galaxies. The cluster  ETG red sequence is  evident, as
well as  the larger  scatter in  color of the  field ETG  around that
sequence.

\begin{figure*}
\epsscale{1.}
\includegraphics[scale=.35]{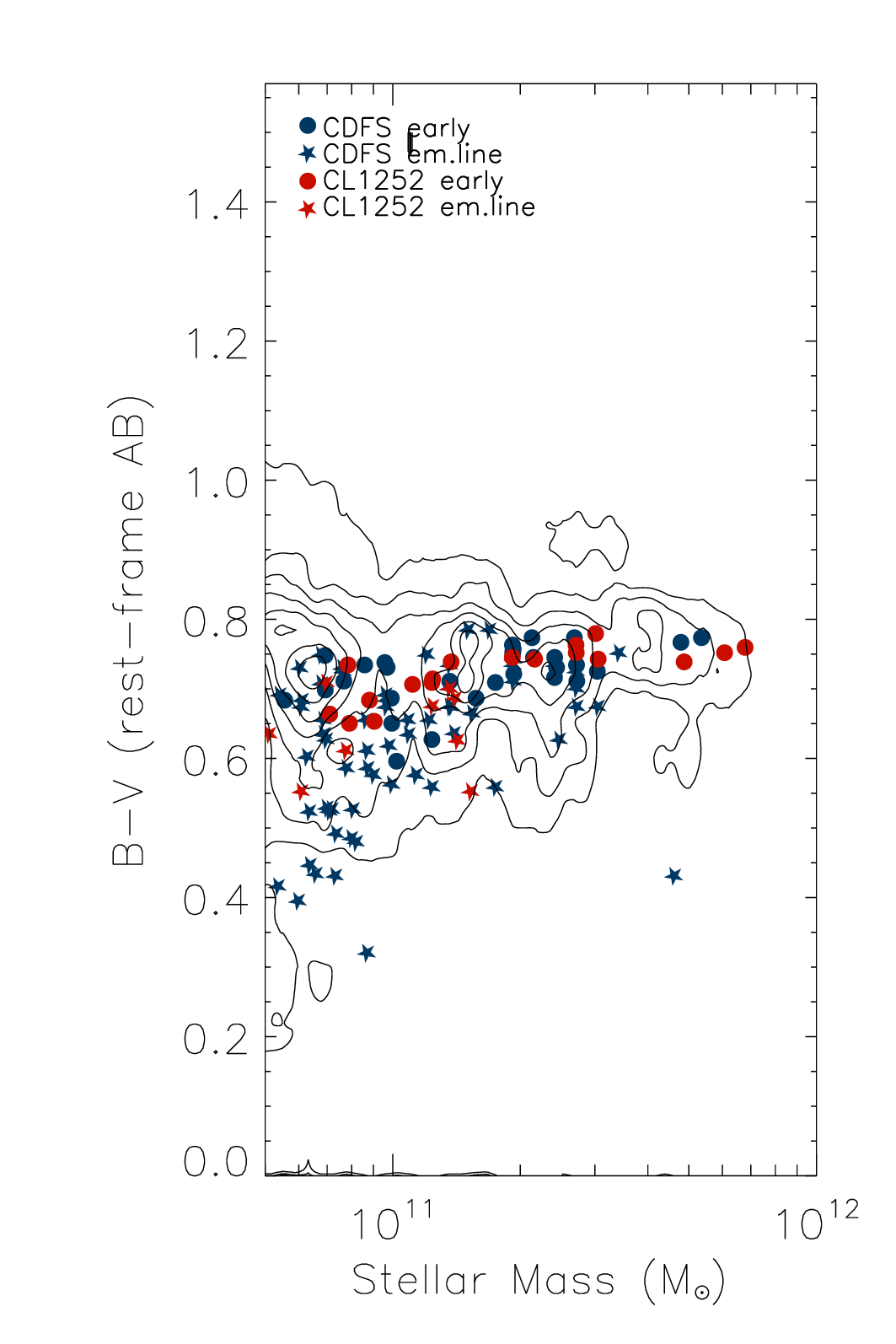}
\includegraphics[scale=.35]{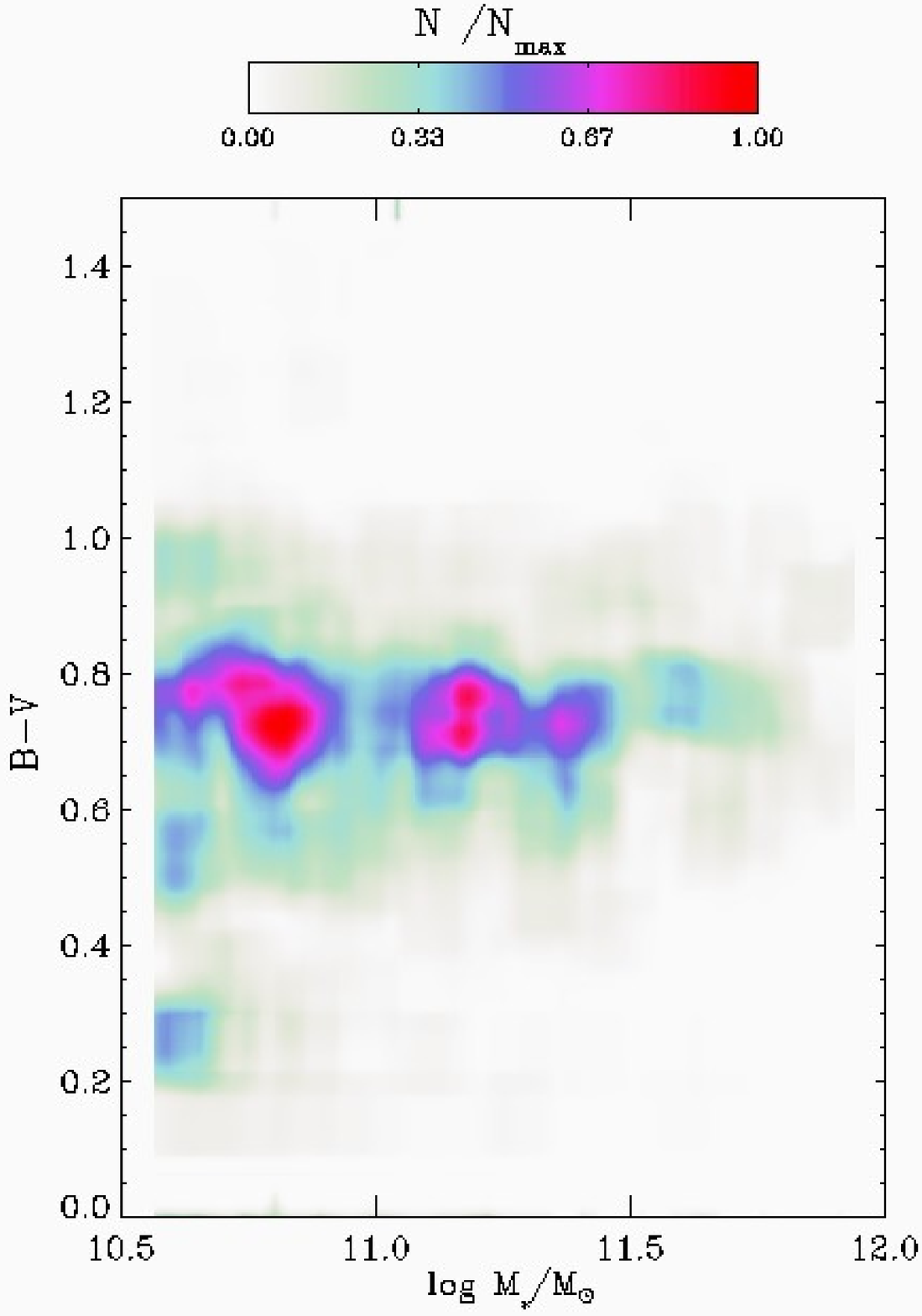}
\caption{$B-V$  Color  versus Stellar  Mass  diagram  of the  observed
  samples of cluster and field galaxies ({\it left panel}, symbols are
  the  same  as  in  Fig.~\ref{UMB_mass})  with  over-plotted  contours
  obtained with the models  of \citet{Menci08} ({\it right panel}) for
  galaxies  in clusters; the  color code  represents the  abundance of
  galaxies in a given ($\frac{log M_{*}}{M_{\odot}}-(B-V)$) bin.}
\label{BMV_mass_theory}
\end{figure*}

In Fig.~\ref{BMV_mass_theory} a similar diagram is compared
with the predictions of the \citet{Menci08} semi-analytical models for
galaxies in  clusters (defined  as host Dark  Matter haloes with  $M >
10^{14}  M_{\odot}$);   the  color  code   represents  the  abundance,
normalized to the maximum value, of  galaxies in a given $mass-(B-V)$ bin.

We note  that the color and  scatter of the sequence  predicted by the
models indicate that  the existence of ETG confined  to a narrow CMR
by   $z\approx  1.2$   is  indeed   consistent  with   predictions  of
hierarchical models including AGN feedback.  However, the latter still
yield  a somewhat  flatter slope  of  the CMR  and an  excess of  red,
low-mass galaxies.   These discrepancies  constitute a
common  feature  of all  hierarchical  models,  due  to the  following
physical processes:  i) the biasing effect,  causing low-mass galaxies
residing  in high-density  environment  to collapse  earlier; ii)  the
starbursts,  present  mainly  at  high-redshifts in  biased  density
environments (like  those originating the  clusters), triggering early
star  formation at  $z\gtrsim 2$;  iii) the  ''strangulation'' effect,
namely, the stripping of gas  in galaxies with shallow potential wells
(such  gas is  included in  the  intra-cluster medium).   In fact,  in
hierarchical models, low-mass galaxies are the main cause of the
larger fraction of red objects characterizing the galaxy population in
high-density environments.

\section{Results and Discussion}

\subsection{The dependence of ETG scaling relations on environment}

The FP  is known to be a  powerful tool for studying  the evolution of
ETGs \citep{Djorgo87}.  In as similar  way to the small scatter of the
color-magnitude  relation  \citep{Bower92}, the  tightness  of the  FP
\citep{Jorgensen96,  Bernardi03}  constrains  the homogeneity  of  the
ETG  stellar population.   Because of its  dependence on
galaxy  luminosity,  the FP  is  sensitive  to  recent star  formation
episodes.

One of its projections shows  a tight relation between  the effective
radius, $R_{e}$, and the  mean surface brightness $<\mu_{e}>$ measured
inside  $R_{e}$, also  known  as  the \citet[][hereafter  KR]{kor77}
relation:
\begin{equation}
<\mu_{e}> = \alpha + \beta log (R_{e}) ,
\end{equation}
where  the slope  $\beta  \simeq 3$  is  found to  be  costant out  to
$z\simeq  0.65$  \citep{LaBarbera03},  while  the  value  of  $\alpha$
depends  on the  photometric  band adopted  to  derive the  structural
parameters.   Here  we   adopt  the  KR  as  one   of  the  tools  for
investigating  the  structural  properties   of  ETG  with  the  aim  of
understanding the  role of the  environment in shaping ETG  of similar
masses and optical-to-infrared colors.

\begin{figure*}
\epsscale{1.}
\plotone{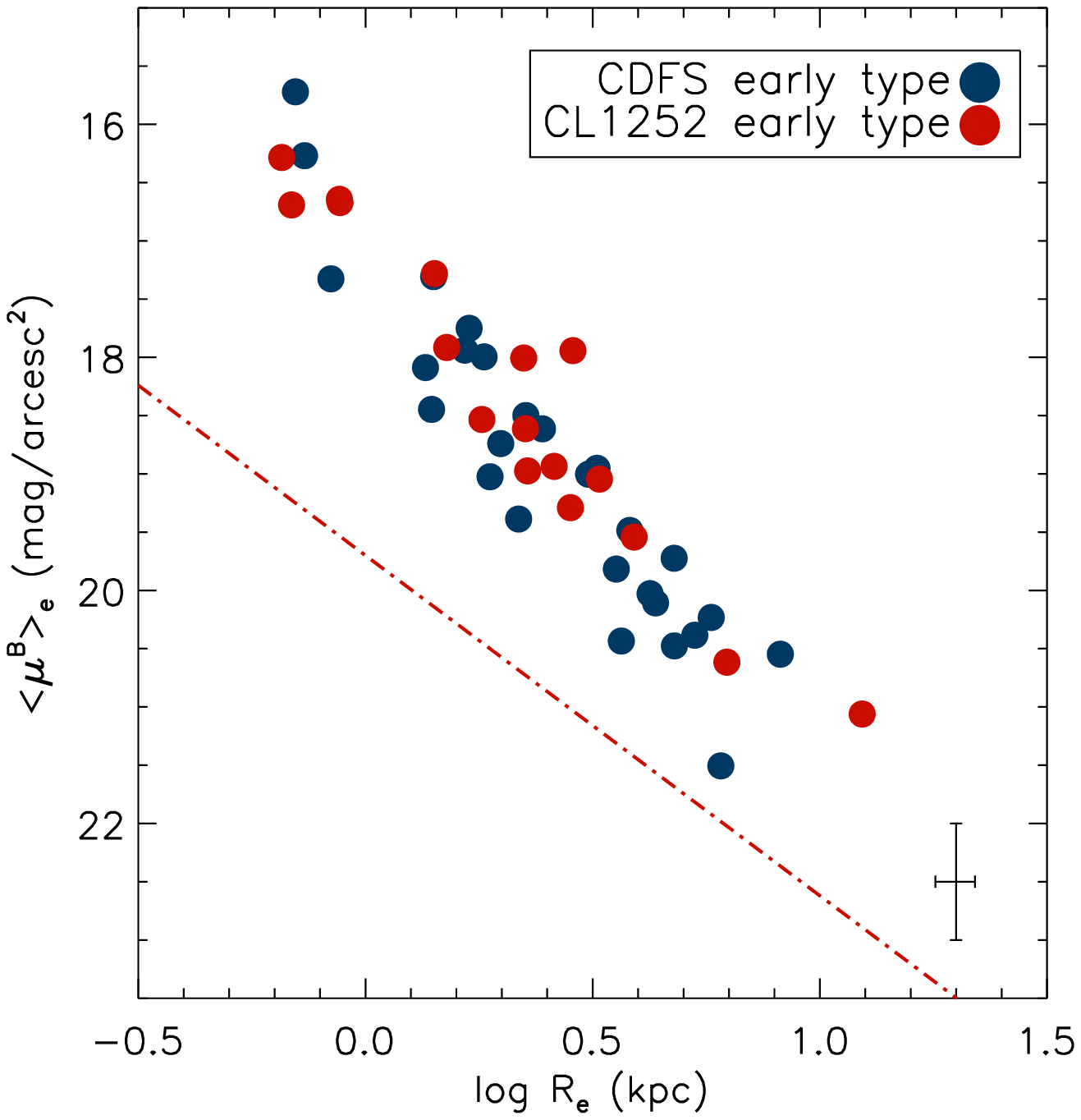}
\caption{Mean surface brightness  $<\mu_{e}>$ versus effective radius,
  $R_{e} [kpc]$. The Kormendy  relation in the rest-frame $B$-band for
  our  ETG in  the field  (filled blue  circles) and  in  the cluster
  (filled  red   circles).  All  the   data  are  corrected   for  the
  cosmological  dimming   $(1+z)^{4}$.  The  red   dotted-dashed  line
  represents  the  KR  at  $z  \sim 0$  found  by  \citet{LaBarbera03},
  K-corrected  to  our  rest-frame  $B$-band.  The error  bar  in  the
  bottom-right is  representative of the typical  uncertainties of our
  measurements.}
\label{Kormendy}
\end{figure*}

In Fig.~\ref{Kormendy}, we find very  similar KRs for the two samples.
Both  the derived  zero points  and slopes  are consistent  within the
errors.   These relations  show that  at the  effective  radius, large
(massive) galaxies  are fainter than small galaxies  regardless of the
environment.   This in  turn indicates  that large  galaxies  are less
dense than small  galaxies in both the cluster and  field at $z \simeq
1.2$.  For comparison,  we overplot  the  KR at  $z \sim  0$ found  by
\citet{LaBarbera03}  (dotted-dashed  red  line),  K-corrected  to  our
rest-frame $B$-band.  Our galaxies are  brighter by 1-2mag than at low
redshift, a discrepancy that other studies at $1.0 \lesssim z \lesssim
1.4$ have  also found  difficult to explain  with the  assumption that
galaxies undergo only a pure luminosity evolution with redshift (e.g.,
\citet{Longhetti07}).   In fact,  our  galaxies show  an evolution  of
$<\mu_{e}>$ which exceeds $\sim 2$  times the one expected in the case
of   pure   luminosity   evolution   ($\sim  1$mag).    According   to
Eq.~\ref{eq:mu_e},  the other  quantity affecting  $<\mu_{e}>$  is the
effective radius.  Therefore to recover this discrepancy we can assume
that, as a function of redshift, ETG undergo a size evolution as well:
the effective radius  of ETG should increase by  at least factor $\sim
1.5$ from  z$\simeq 1.2$ to z$\sim 0$  both in the cluster  and in the
field environment.

\begin{figure*}
\epsscale{1.}
\plotone{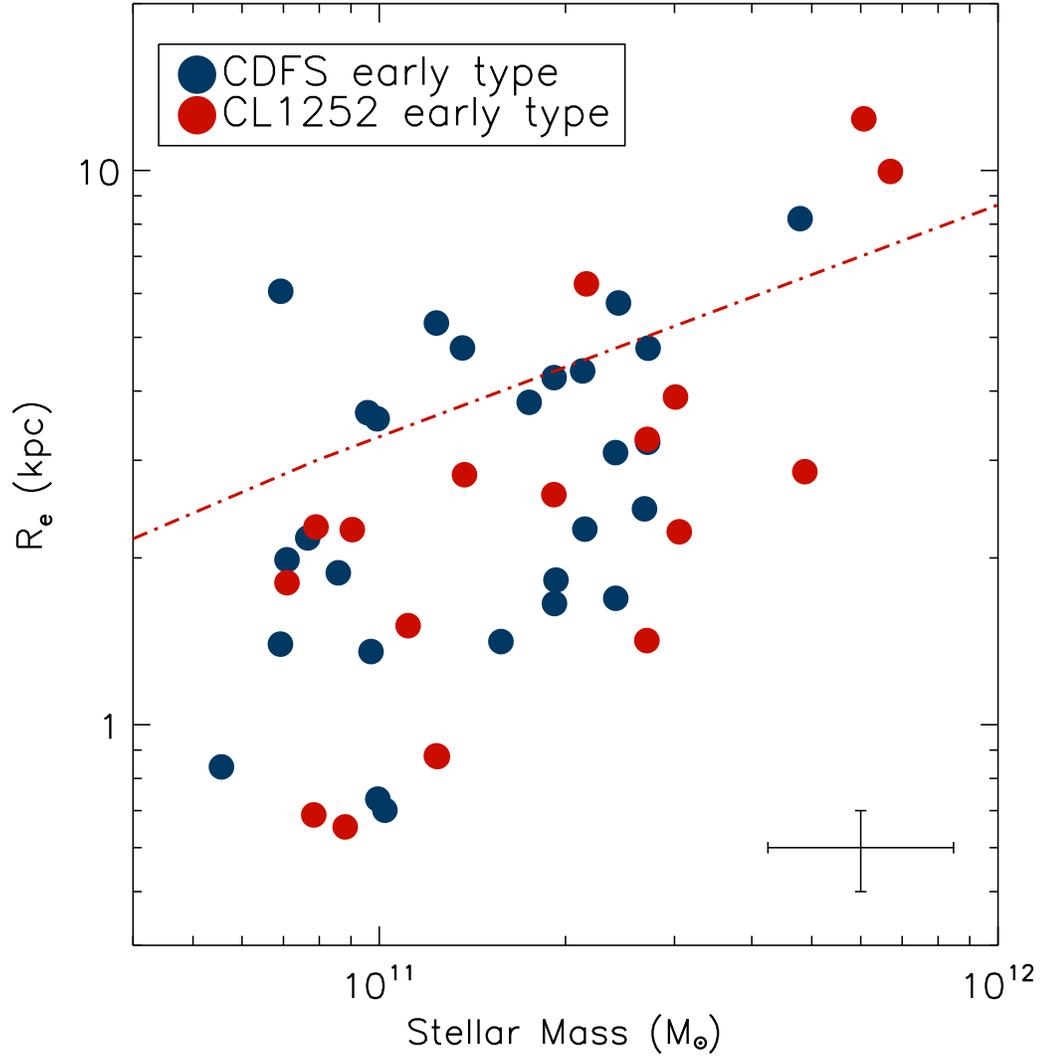}
\caption{The  stellar mass  vs.  size  relation  for the  ETG in  the
  cluster  (filled  red  circles)   and  in  the  field  (filled  blue
  circles). The red dotted-dashed  line represent the same relation at
  $z \sim  0$ found  by \citep{Shen03} with  SDSS data. The  mean size
  relative error is $<$ 20\%.  Uncertainties in stellar mass are $\sim
  0.15$ dex.}
\label{mass_size}
\end{figure*}

Recent  studies of  the  dependence  on environment  of  the size  vs.
stellar  mass   relation  \citep{Trujillo04,  McIntosh05,  Trujillo06,
  Trujillo07} found that the  bulk of galaxies with comparable stellar
masses to  ours were at least  a factor 2 smaller  at higher redshifts
than  locally.  This  is  qualitatively consistent  with the  observed
trend  in our  data,  (see Fig.~\ref{mass_size})  when  the sizes  and
masses our  samples are  compared with the  local relation for  ETG in
SDSS \citep[dotted-dashed red line;][]{Shen03}.  We find no-dependence
on the environment of the $R_{e}$ vs.  $M_{*}$ relation, implying that
cluster  and  field ETGs  must  undergo  similar  luminosity and  size
evolution  to match  the typical  values found  for the  ETG  at lower
redshifts.  To  explain how compact  galaxies observed at  $z>1$ could
possibly  end-up  on  the  local  relation,  a  possible  evolutionary
mechanism  that grows  stellar mass  and  size has  been suggested:  a
dissipationless   (``dry'')  merging   of   gas-poor  systems   (e.g.,
\citep{Ciotti01, Nipoti03, Khochfar03,  Boylan06} that is efficient in
increasing  the size of  the objects,  while remaining  inefficient at
forming new stars.

In   the  local  universe,   the  SFR   per  stellar   mass  (specific
star-formation   rate,   SSFR)   correlates   strongly   with   galaxy
concentration, effective  radius and the average  surface stellar mass
density \citep[$\sigma_{50}$;][]{Kauffmann03,Brinchmann04}. A striking
similarity of  cluster and  field galaxies at  $z \simeq$1.2  is again
shown in  Fig.~\ref{andrew} where we plot, $\sigma_{50}$,
\begin{equation}
\sigma_{50} = \frac{0.5 M_{\star}}{\pi R_{e}^{2}} ,
\end{equation}
versus the  stellar mass,  and compare them  with similar data  in the
literature  drawn  from \citet{Zirm07}  at  $z  \sim 2.5$.   Quiescent
Distant Red  Galaxies (qDRGs) are drawn as  red ellipses, star-forming
DRGs (sDRGs) are  drawn as open red stars,  while blue stars indicates
Lyman Break  Galaxies (LBGs)  from the same  work.  While some  of our
galaxies are almost as  dense as the \citet{Zirm07} and \citep{Toft07}
quiescent  distant  red galaxies  (qDRGs),  both  our samples  (filled
circles) overlap the region occupied by other $1.0 \lesssim z \lesssim
1.5$  galaxy  samples  \citep{Trujillo06, Daddi05,  vdW06,  Rettura06}
(open red  squares, open red  circles, open black circles,  open black
squares, respectively).   As a comparison  we also overplot  the local
relation for ETG in SDSS (red dotted-dashed line) calculated from the
mass-size   relation  of   \citet{Shen03}.   It   is  very   clear  in
Fig.~\ref{andrew} that  the bulk of  our galaxies in both  samples have
much larger  densities that their local counterparts.   To account for
this  effect  in  the  context  of  a  plausible  formation  scenario,
semi-analytical modeling  (e.g. \citet{Khochfar06}) suggests  that ETGs
formed in gas-rich mergers can  result in very dense stellar cores, as
the gas is  driven to the center of  this ``wet'' (dissipative) merger
where it  very efficiently produces massive  starburts.  Galaxies that
merge in the  early universe are likely to  be gas-rich.  Consequently
the dense  nature of this objects  could be the result  of much denser
conditions of  the universe  at the time  of their formation.  We note
that  our   finding that there is no dependence  on  the   environment  of  the
$\sigma_{50}$  vs,  $M_{*}$  relation  at z$\simeq  1.2$  can  provide an
important datum for models of galaxy formation.

%Figure Sigma vs. Radius
%We also find  a trend for $z\simeq$1.2 cluster  galaxies of increasing
%$\sigma_{50}$ as a function of the increasing distance from the center
%of the cluster (see Fig.~\ref{distanza}).   We can interpret this as a
%direct consequence of  the . This effect causes  the more massive ETGs
%to be  more likely found in the  center of the clusters  than in their
%outskirts.  According to  the KR, at the effective  radii, these large
%galaxies will be fainter than the smaller, more compact, galaxies that
%live in  the outer  regions.  Therefore less  dense galaxies  are more
%likely to  be found in cluster  cores , with the  average surface mass
%density thus increasing with the projected distance from the center.

\subsection{The dependence of ETG ages and star formation histories
  on their environment}

As we  apply the method  described in section  \S 3.2, we are  able to
directly compare  the distribution of star-formation  weighted ages in
the  field  and  in  the  cluster.   As shown  in  the  top  panel  of
Fig.~\ref{timing},  we  find  the  overall  relative  distribution  of
cluster and field ETG ages to be very similar.  This result implies no
significant delay  in relative  formation epochs is  found for  ETG in
either environments.   We find that  $\sim 80$\% of massive  ETGs have
ages in the range $3.5 \pm 1.0$ Gyr in both cluster and field.

To investigate  the  dependence of  this  result on  the
actual  stellar  population synthesis  code  adopted,  we compare  our
current results (based on BC03 models) with those obtained with a set
of  similar  dust-free CSPs  templates  built with  \citet{Maraston05}
models,   adopting   the    same   exponentially-declining   SFHs   of
Eq.~\ref{eq:sfhs},     and    assuming    solar     metallicity    and
\citet{Salpeter55}  IMF.  The  result  of the  analysis  based on  M05
models  is shown  in the  bottom panel  of Fig.~\ref{timing}  where we
still find the cluster and field relative age distributions to be very
similar, despite the fact that the contribution of the TP-AGB stars in
these models  are implemented  in a different  way.  However,  here we
find that $\sim 60$\% of galaxies have ages in the range $3.5 \pm 1.0$
Gyr; M05 models favor slightly younger ages ($\sim1-2$ Gyrs) for $\sim
$20\% of ETG  in both environments.  This effect  can be explained by
the fact that, at about 1  Gyr, M05 models account for a larger amount
of  2 $\mu m$  flux than  BC03 models  of the  same age,  resulting in
significantly redder  color at  younger ages, thus  can favor  $\overline t
\sim$ 1-2Gyrs best-fits in a few cases.

%implementing  a different treatment of,
%Hence  we note that  estimates of the  star-formation weighted
%ages  of $z  \sim 1$  ETGs, if  based on  such an  extended wavelength
%baseline  (8-9 bands  from rest-frame  FUV  to NIR),  are not  largely
%affected by stellar population synthesis model discrepancies.

To  summarize,  we  find   that,  regardless  of  the  actual  stellar
population synthesis code adopted, cluster galaxies ages have the same
relative  distribution as their  field contemporaries:  no significant
delay in  their formation  epochs is found,  within the  errors ($\sim
0.5Gyr$).   This result  is  at  variance with  some  versions of  the
hierarchical    model    of    galaxy    formation    and    evolution
\citep{Diaferio01,   Delucia06}  and   with   fossil  record   studies
\citep{Thomas05, Clemens06}, but in remarkably good agreement with the
ones  derived  by   \citet{vandokkum07}  and  \citet{dSA06}  from  the
evolution of the $M/L$ ratio.  It should be noted that similar results
are found by other works using independent methods and data-sets.

In the top-left panel of Fig.~\ref{quartetto} we plot for both samples
each galaxy  age, $\overline  t$, as a  function of stellar  mass.  We
note  that  the  age  of   ETG  increases  with  galaxy  mass  in  all
environments,  which   is  in  agreement  with   the  so--called  {\it
  downsizing}  scenario  of  galaxy formation  \citep{Cowie96}.   This
effect   can  also   be   seen   in  the   the   top-right  panel   of
Fig.~\ref{quartetto},  where we  plot  our galaxies'  lookback time  to
formation as  a function of  their stellar mass, in  both environments.
Our result  is in  agreement with the  one obtained with  an
independent method  and data-set by \citep{dSA06} and  based on z$\sim
1$ ETG ages estimated from the FP parameters (see their Fig. 3).

Despite of the fact that cluster and field galaxy formation epochs are
found to be similar, it could still be possible that the timescales of
their SFH  are significantly different.   Firstly, the data  show that
the distribution of cluster  and field optical colors is significantly
different. As  a function  of the stellar  mass, cluster  galaxies are
found to  lie on a very  tight red-sequence, while those  in the field
populate     the    color-sequence     with    a     larger    scatter
(Fig.~\ref{UMB_mass}). Secondly,  in \citet{Gobat08} we  find that the
averaged  spectrum  of the  cluster  galaxies  has  a more  pronounced
4000\AA \mbox{} break than that of the field sample.

Both these pieces of observational evidence find a natural explanation
in the framework of our  modeling.  As shown in the bottom-right panel
of Fig.~\ref{quartetto}, as  a function of stellar mass,  we find that
field  ETG  span a  larger  range  of  timescales than  their  cluster
contemporaries, which are formed with the shortest $\tau$ at any given
mass.   According  to  our  models,  cluster ETG  are  found  to  have
experienced more  similar star-formation  histories.  As shown  in the
bottom-left  panel   of  Fig.~\ref{quartetto},  cluster   ETG  form  a
color-age sequence with much smaller scatter than the field ones.

As  discussed in  \S 2.1,  we  recall that  our field  sample is  more
deficient in lower mass objects than the cluster sample because of the
different depths of  spectroscopic observations. However, even
  if  the field  sample were  corrected for  completeness,  this would
  likely result in a larger fraction of field ETG at low mass,
  which are  the ones that we found with  longer $\tau$. Hence
  this would amplify the  difference between the typical timescales of
  the two samples, and so not affect our conclusions.

\subsection{The dependence of ETG FUV magnitudes on their environment}

The rest-frame FUV ($\sim \lambda ~  1700 \AA$) SED is a crucial range
where hot ($> 9000 K$),  massive ($M > 2 M_{\odot}$), short-living ($<
1Gyr$) stars emit most of their light.  Thus it is a wavelength domain
which is very sesnsitive to current or recent star formation.  Most of
the light from  ETG is emitted in the optical  and the NIR rest-frame.
However, the FUV can be used  as a good tracer of the residual current
star-formation  and to  trace  back,  within the  last  Gyr, the  most
recent episode of star-formation. About 100Myr after star formation
ceases,  an ETG  spectrum  becomes dimmer  and  redder. Therefore,  the
fainter the  rest-frame UV emission  is, the earlier  the
star  formation must  have stopped.   Over time,  the  galaxy spectrum
fades  and  slowly reddens  as  the  4000  $\AA$ break  becomes  more
pronounced.   Here we have used the  {\it  VLT}/VIMOS  $U$-band
observations  described  in  \S   2.2  to  empirically  constrain  the
dependence  on  the environment  of  the  most  recent star  formation
processes in z$\simeq 1.2$ ETGs.

However,  when analyzing  UV rest-frame  fluxes of  massive ETG  it is
important to recall  that core helium burning stars  on the horizontal
branch (HB) are known to  produce a ``UV upturn'' feature \citep{Yi97,
  Yi99}.    This   effect    can,   in   principle,   complicate   the
disentanglement of the contributions to the UV spectrum of the evolved
and young stellar populations.  However, the onset of the HB typically
takes 9  Gyr, meaning that, by  z=1.2 (when the Universe  is only 5Gyr
old), not enough cosmic time  has elapsed for this population of stars
to appear.   Hence the UV flux  seen in our sample  ETG must originate
only from young stars.

In Fig.~\ref{uband} we show  $U$-band magnitudes (1'' radius aperture;
rest-frame $\sim  1700 \AA $)  as a function  of stellar mass  for the
ETG  detected  in  the  field  (filled blue  circles).   Solid  lines
represent the  1$\sigma$ limit magnitudes  of both data-sets  (in blue
for  CDFS,  in  red  CL1252).  Dashed lines  represent  the  3$\sigma$
limits.  As already  pointed  out  in \S2.2,  the  combined effect  of
shorter  total exposure times  and higher  galactic extinction  at the
location  of CL1252,  directly translates  into a  $\sim  1mag$ deeper
$U$-band photometry  for the  CDFS.
% The 3-$\sigma$  depth for  CDFS is
%$U$=28.27, while the corresponding depth for CL1252 is $U$=27.3.

A  large fraction (75  \%) of  field ETG  are ($>3 \sigma$)
detected  in the  deep CDFS  images.   The observed  magnitude of  the
median stack of these detections is $U$=27.46 mag (blue dotted line of
Fig.~\ref{uband}),   corresponding   to   a  SFR=0.47   $M_{\odot}/yr$
\citep{Sawicki06}.  An image of the  median $U$-band stack of the ETGs
detected  in the  field, is  displayed  in the  bottom-left corner  of
Fig.~\ref{uband}.

Since none of the CL1252 ETG is actually detected in our $U$-band data,
we use their median stack, shown in the middle of Fig.~\ref{uband}, to
provide a robust  upper limit of $U>27.3$ mag  for the CL1252 early-type
population, which corresponds to SFR$< 0.55 M_{\odot}/yr$.

We  note that  the non  detection  of the  CL1252 ETG  cannot only  be
attributed to the  shallower $U$-band data for the  cluster.  To prove
this we have  simulated how the CDFS detected ETG  would appear in the
1252  data  (more  details  can  be found  in  \citet{Nonino08}).   We
randomly placed the  20 $U$-band detected CDFS ETG  (0.4 mag dimmed to
match the  relative difference in  galactic extinction) in  the CL1252
maximally exposed  region, avoiding objects detected  in the $U$-band,
and repeated this  step 30 times. Hence we  generated 32 median stacks
of  18 simulated  galaxies  each,  picking up  at  random amongst  the
cutouts.   Aperture photometry  (1'' radius)  on these  simulated {\it
  CDFS@1252} stacks results in a median value of $U$=27.8 mag which is
in agreeement with the dimmed inputs of the simulations.
%which  corresponds  to  a  SFR=$0.34 M_{\odot}/yr$.  
In the  bottom-right corner of  Fig.~\ref{uband} we show one  of these
stacks, which would be clearly  detected in 1252.  Comparing this last
value to the upper limit we measured in the cluster data, we can state
with confidence that cluster ETG are intrinsically fainter by at least
0.5 mag  in the observed  $U$-band than their field  contemporaries of
similar mass and optical-to-infrared colors.

This observational  evidence corroborates  the results of  our stellar
population  synthesis analysis described  in the  previous subsection.
In our proposed  scenario, a generally shorter {\it  timescale} of the
star  formation  process among  the  cluster  galaxy population  would
naturally result  in a generally fainter  observed $U$-band magnitudes
compared to the field population at z$\simeq 1.2$.

\section {Conclusions}

We   have  obtained   photometric  parameters: PSF-matched  aperture
magnitudes in  9 bands from  FUV to NIR rest-frame;  and morphological
information: effective radius,  average surface  brightness, S\'ersic
index and average surface stellar mass density for mass-selected samples
of  45 cluster and  field massive  ($M >  5 \cdot  10^{10} M_{\odot}$)
early-type galaxies  at z$\simeq  1.2$.  Apart from  a lower  level of
spectroscopic completeness for the  least massive field galaxies, that
we find not  to affect our conclusions, our  sample has the advantage
of being photometrically complete at  our mass limit and having galaxy
types assigned spectroscopically.

For both samples  we also have derived stellar  masses, ages, and star
formation   histories,  parameterized   as   timescales,  $\tau$,   of
exponentially declining  CSP model templates  built with BC03  and M05
models.

From the  data analysis  performed in this  work, we have  obtained the
following results:

\begin{itemize}
\item{}  Cluster and  field  ETG  lie on  a  similarly tight  Kormendy
  Relation at z$\simeq  1.2$. When compared to the  local relation, our
  galaxies are 1-2mag  brighter than at z$\sim$0, similar  to what has
  been found  by other  studies at  $z \sim 1.0$  in the  field (e.g.,
  \citep{Longhetti07}).  The  evolution of the KR  cannot be explained
  as pure luminosity evolution and we  conclude that ETG must undergo a
  similar size evolution in both environments.
\end{itemize}

\begin{itemize}
\item{} We find no dependence on the environment of the {\it size vs.
    stellar mass}  relation,or for the {\it  average surface
    stellar mass density vs.  stellar mass} relation at z$\simeq 1.2$.
  As a comparison we constrast both of them with the local relations for
  ETG found  in SDSS.   We find that  the bulk  of our ETG  in both
  samples have  much smaller sizes  (and larger densities)  than their
  local counterparts.  Our data therefore indicate a strong size
  evolution for both the cluster and field galaxies.
\end{itemize}

\begin{itemize}
\item{}  We find  no  significant  delay in  the  formation epochs  of
  massive ETG  observed in  the cluster and  in the field  at z$\simeq
  1.2$.   This  result  is  true  robust for  models  that  treat  the
  contribution of the TP-AGB stars in very differnt way (i.e., BC03 or
  M05).   However our  result is  at  variance with  some versions  of
  hierarchical  models \citep{Diaferio01,  Delucia06} and  with fossil
  record  studies \citep{Thomas05,  Clemens06}, but  is  in remarkably
  good agreement with  those obtained from the evolution  of the $M/L$
  ratio.
\end{itemize}

\begin{itemize}
\item{} The age of ETG  increase with galaxy mass in all environments,
  which is in agreement with  the {\it downsizing} scenario.  The site
  of active star formation must  have shifted from the most massive to
  the  less massive galaxies  as a  function of  the cosmic  time. The
  formation epochs  of ETG only depends  on their mass and  not on the
  environment they live in.
\end{itemize}

\begin{itemize}
\item{} We present new  deep $U$-band observations in the rest-frame
  FUV ($\sim \lambda ~ 1700 \AA$) for both samples ETG.  We detected
  75 \% of the field ETG at z$\simeq 1.2$.  The observed magnitude of
  the  median  stack  of  these  detections  is  $U$=27.46  mag,  which
  corresponds to a SFR=0.47 $M_{\odot}/yr$.
\end{itemize}

\begin{itemize}
\item{} None of  the CL1252 ETG was actually  detected in our shallower
  (by $\sim$ 1 mag) $U$-band data, but we used their median stack image to
  provide a robust upper limit of $U>27.3$ mag for the CL1252 early-type
  population, which corresponds to SFR$< 0.55 M_{\odot}/yr$.
\end{itemize}

\begin{itemize}
\item{} Using simulations, we find that the median
  stack of the CDFS ETG could  be clearly detected in the actual CL1252
  data at  $U$=27.8 mag.  Comparing  this value to
  the upper  limit we  measured from the  cluster data, we  can firmly
  state that cluster ETG are  intrinsically fainter by at least 0.5
  mag in the $U$-band than  their field contemporaries of similar mass
  and   optical-to-infrared  colors.   This  observational
  evidence implies that  the last episode of star  formation must have
  happened more recently in the  field than in the cluster.
\end{itemize}

\begin{itemize}
\item{} The data  also show two other compelling pieces of evidence that
  cluster  and   field  SFHs  are
  significantly  different:  1) as  a  function of  the  stellar  mass,
  cluster galaxies  are found  to lie on  a very tight  color sequence
  while the  field galaxies  populate it  with a larger  scatter; 2)  in a
  companion paper based on the same data-set, \citet{Gobat08} find
  that  the averaged  spectrum  of  the cluster  galaxies  has a  more
  pronounced 4000\AA \mbox{} break than that of the field sample.
\end{itemize}

\begin{itemize}
\item{} Finally  we have been also  able to explain both
  these pieces of  observational  evidences  in  the framework  of  our  stellar
  population modeling. Field ETG best-fit models span a larger range
  of timescales  than their  cluster contemporaries, which  are formed
  with the shortest $\tau$ at any given mass.
\end{itemize}

While cluster and  field galaxy observed at z$\simeq 1.2$
form at  approximately the same  time, a high density  environment is
able to trigger more rapid  and homogenous SFHs for the ETGs, limiting
the  range  of  possible  star-formation processes.   In  low  density
environments, this effect  must rapidly fade as ETG  undergo a
much broader range of possible star formation histories.  We also note
that this scenario is in very  good agreement with the one proposed by
\citet{Menci08},  based  on  the  latest  rendition  of  semi-analytic
models.

\begin{figure*}
\epsscale{1.}
\plotone{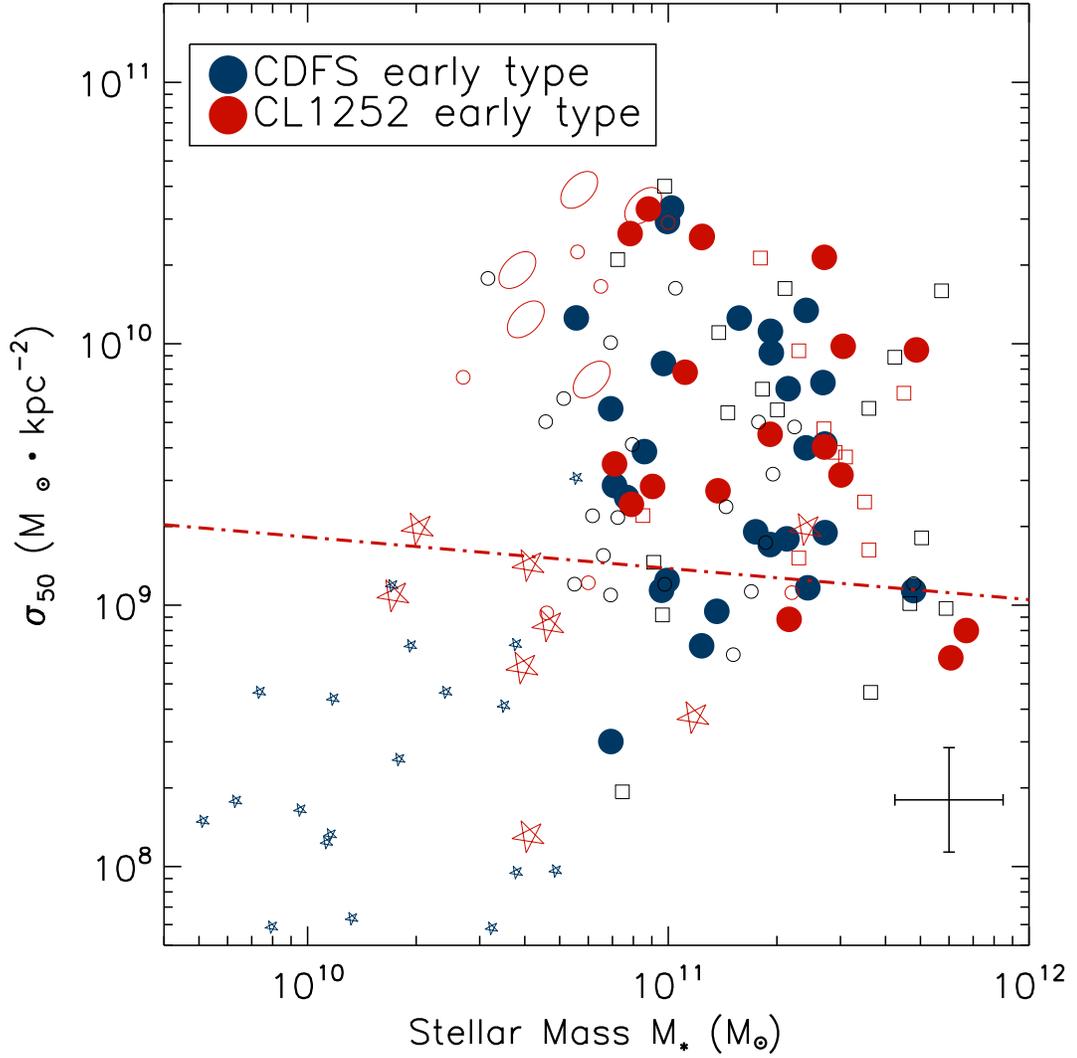}
\caption{The stellar mass vs.  the Average surface mass density within
  the  half-light radius  for  the  ETG in  the  cluster (filled  red
  circles) and in the field  (filled blue circles).  For comparison we
  over-plot $z  \sim 2.5$ quiescent DRGs (qDRGs)  (open red ellipses),
  star-forming DRGs (sDRGs) (open red stars), LBGs (filled blue stars)
  from  \citet{Zirm07}  paper.   Other  samples  of  $1.0  \lesssim  z
  \lesssim   1.5$   ETG   are   also   drawn  from   the   works   of
  \citep{Trujillo06}  (open red  squares),  \citep{Daddi05} (open  red
  circles), \citep{vdW06} (open  black circles), and \citep{Rettura06}
  (open black  squares).  The red dotted-dashed  line represent
  the local  relation for ETG  in SDSS calculated from  the mass-size
  relation  of \citet{Shen03}. The  error bar  in the  bottom-right is
  representative of the typical uncertainties of our measurements.}
\label{andrew}
\end{figure*}

\begin{figure*}
\epsscale{1.0}
\includegraphics[angle=90,scale=.55]{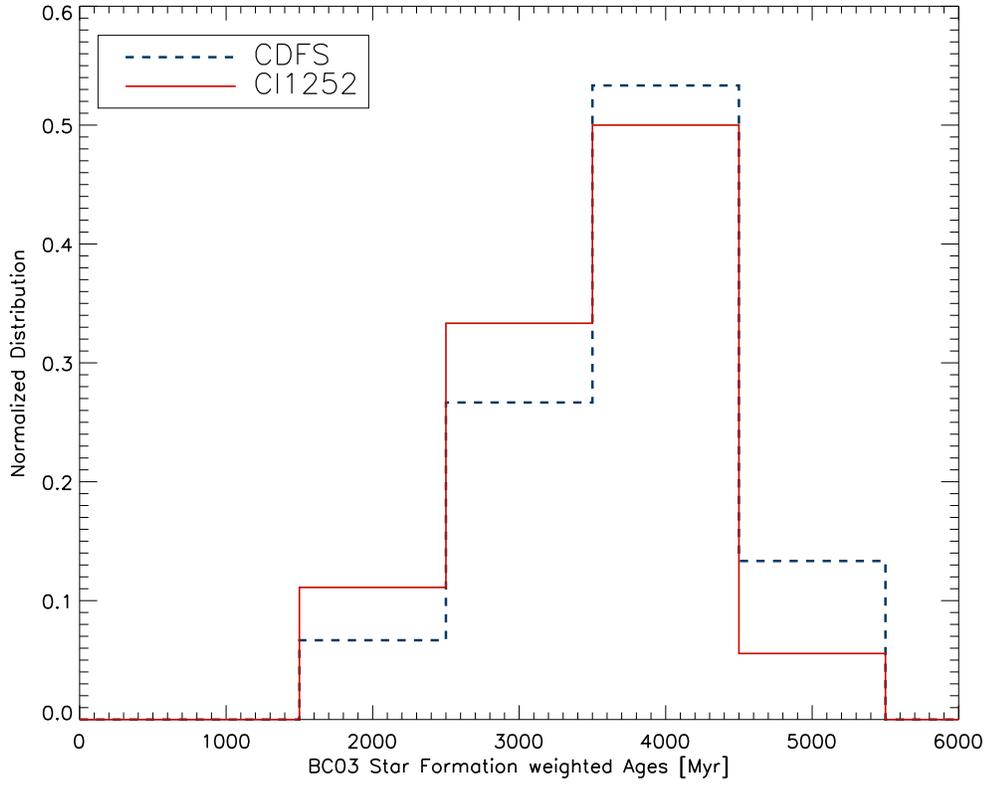}
\includegraphics[angle=90,scale=.55]{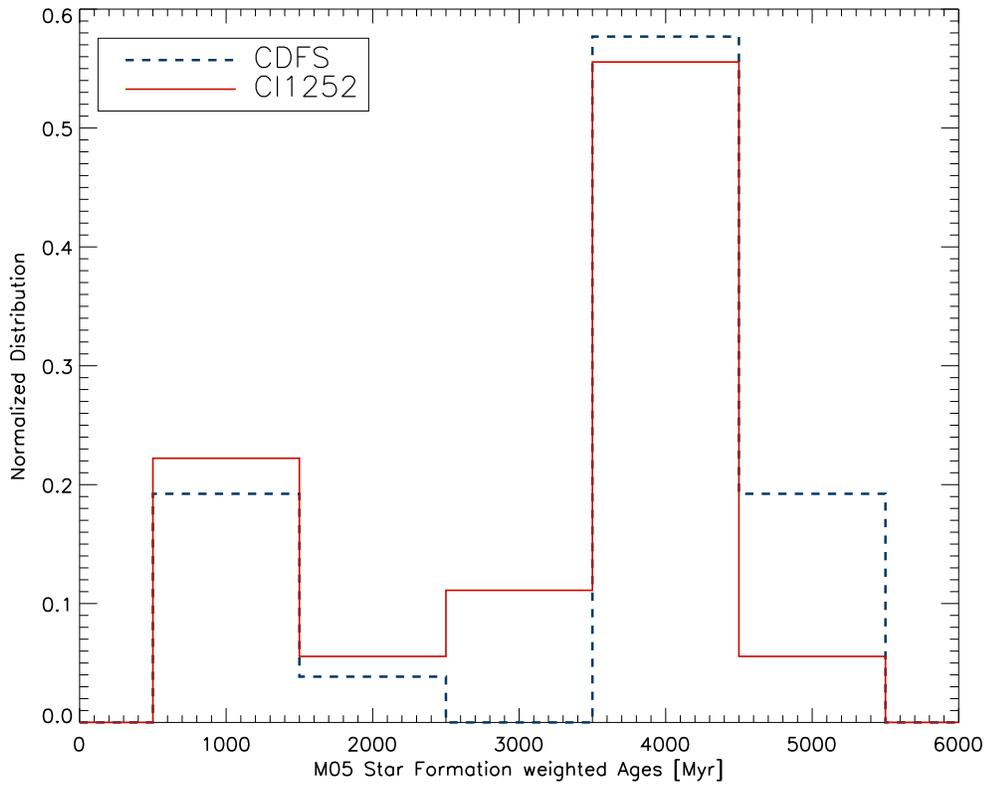}
\caption{{\it Top  panel}: Formation epochs of ETGs:  Histogram of the
  field (dashed blue line) and cluster (solid red line) star-formation
  weighted  ages derived with  the BC03  models.  {\it  Bottom panel}:
  Histogram  of  the field  and  cluster star-formation-weighted  ages
  derived with the M05 models.}
\label{timing}
\end{figure*}

\begin{figure*}
\epsscale{1.0}
\includegraphics[scale=.6]{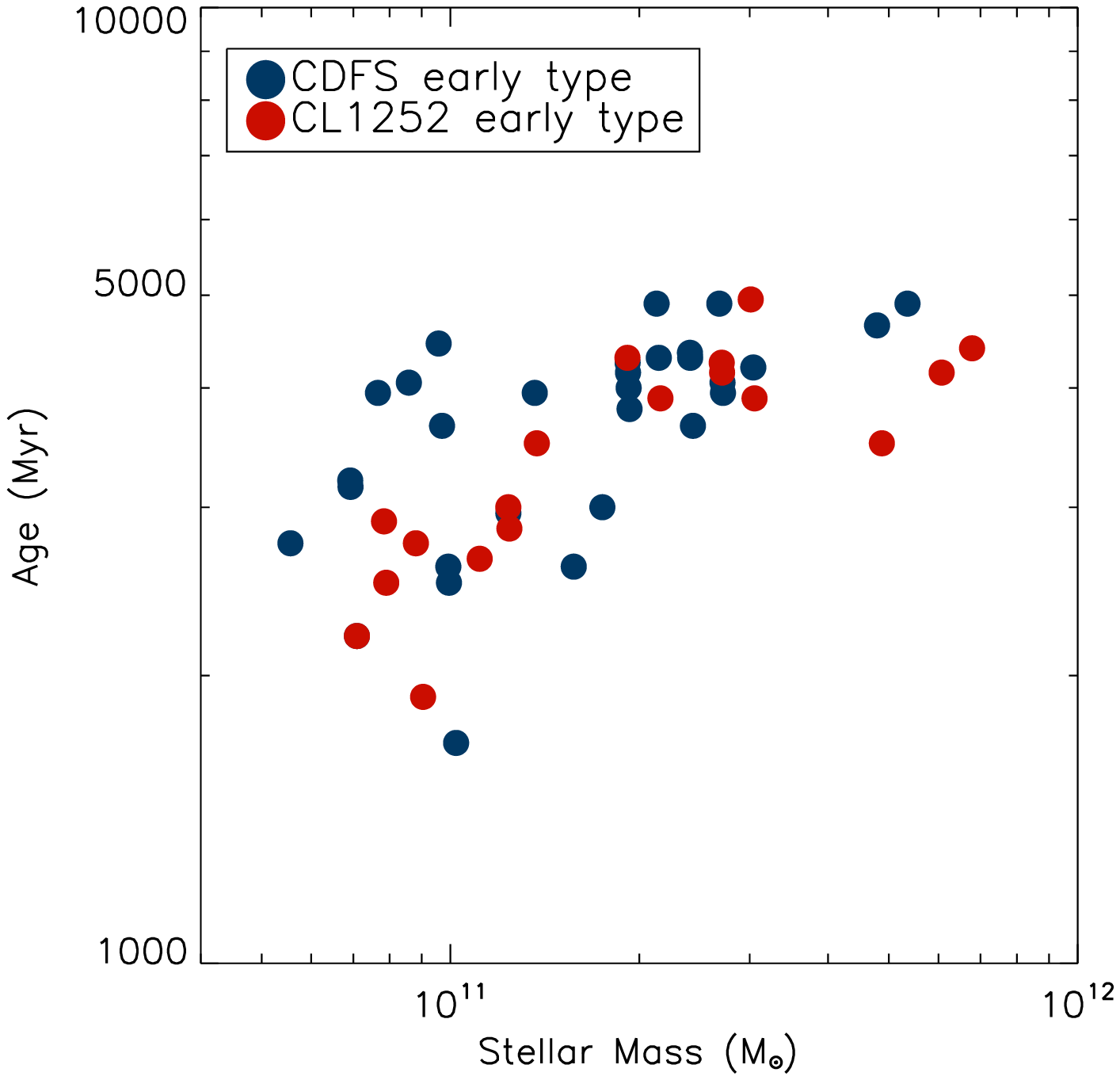}
\includegraphics[scale=.6]{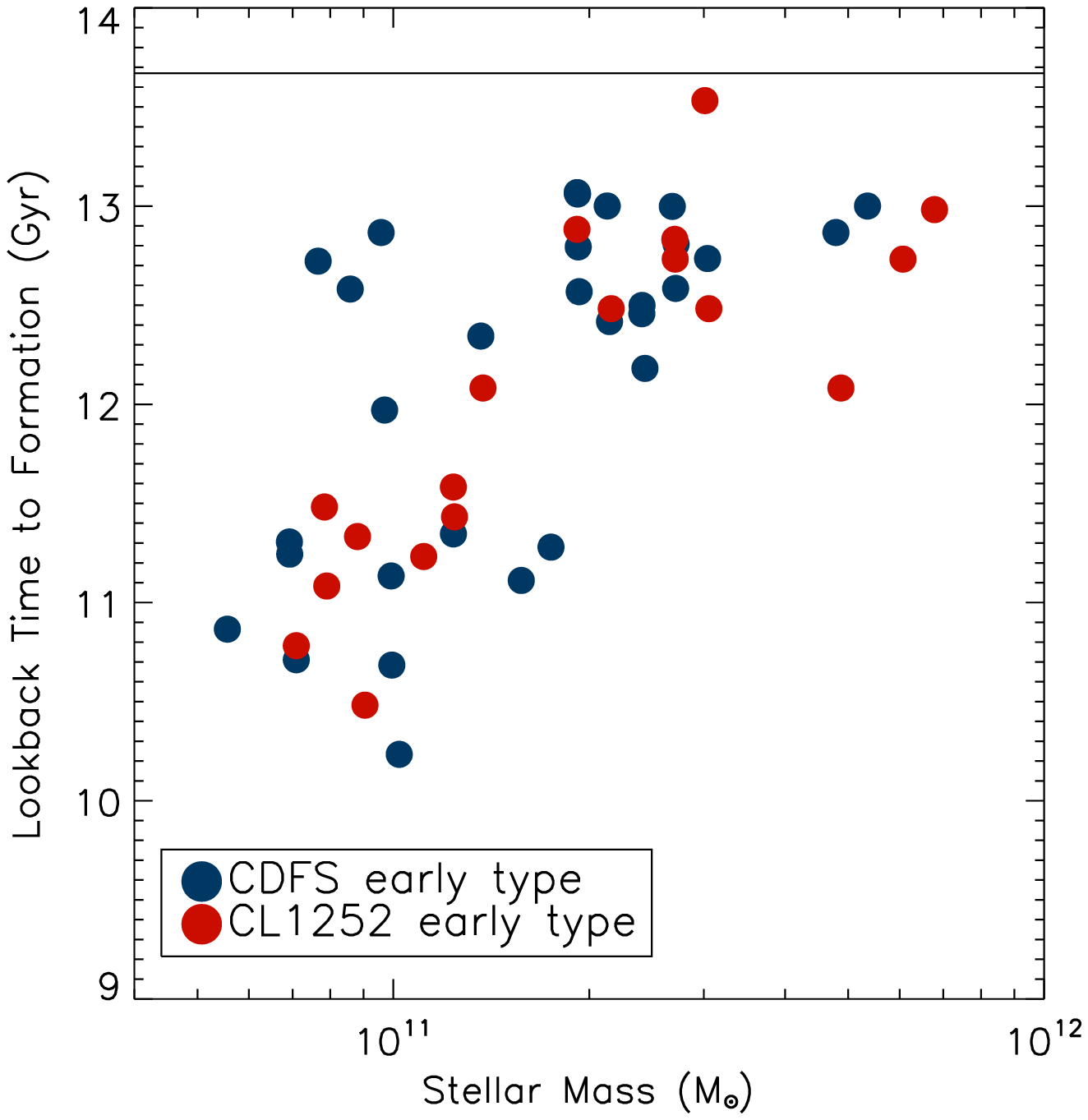}
\includegraphics[scale=.6]{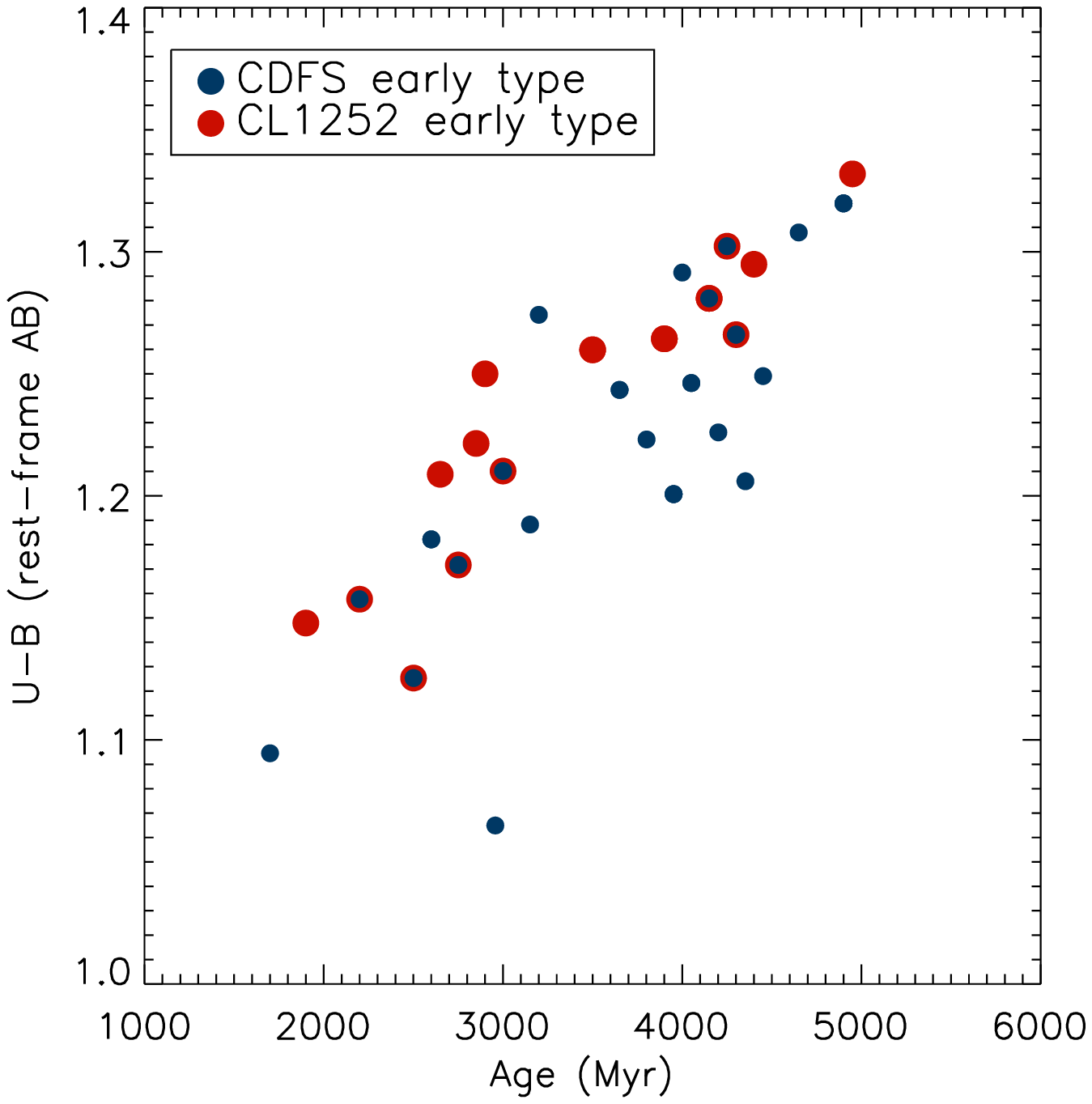}
\includegraphics[scale=.6]{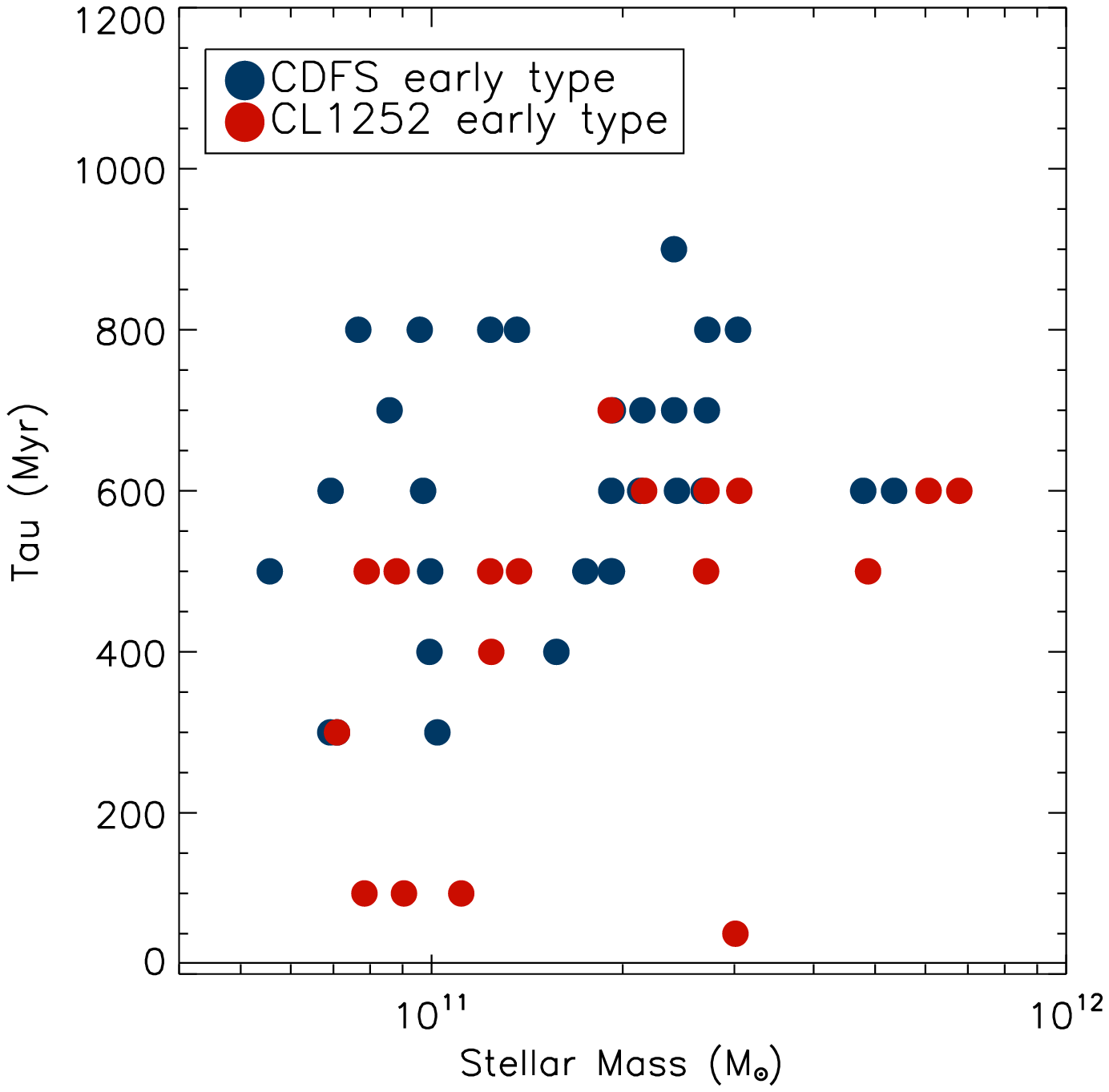}
\caption{{\it  Top-left  panel}:  Stellar  mass  vs.   star  formation
  weighted ages for the ETG in the cluster (filled red circles) and in
  the  field  (filled  blue  circles).   {\it  Top-right  panel}:  The
  dependence of  the lookback  time to formation  on the  galaxy mass.
  {\it Bottom-left panel}:  Rest-frame $U-B$ color vs.  age  of ETG in
  both  environments.    {\it  Bottom-right  panel}:   Formation  {\it
    timescales}, $\tau$, of ETG as a function of their stellar mass in
  both the field  and cluster environment.  The mean  error in stellar
  age  is 0.5  Gyr. Uncertainties  in stellar  masses and  $\tau$ are
  $\sim 0.15$ dex and $\sim 0.2$ Gyr, respectively.}
\label{quartetto}
\end{figure*}

\begin{figure*}
\epsscale{1.0}
\includegraphics[scale=.8]{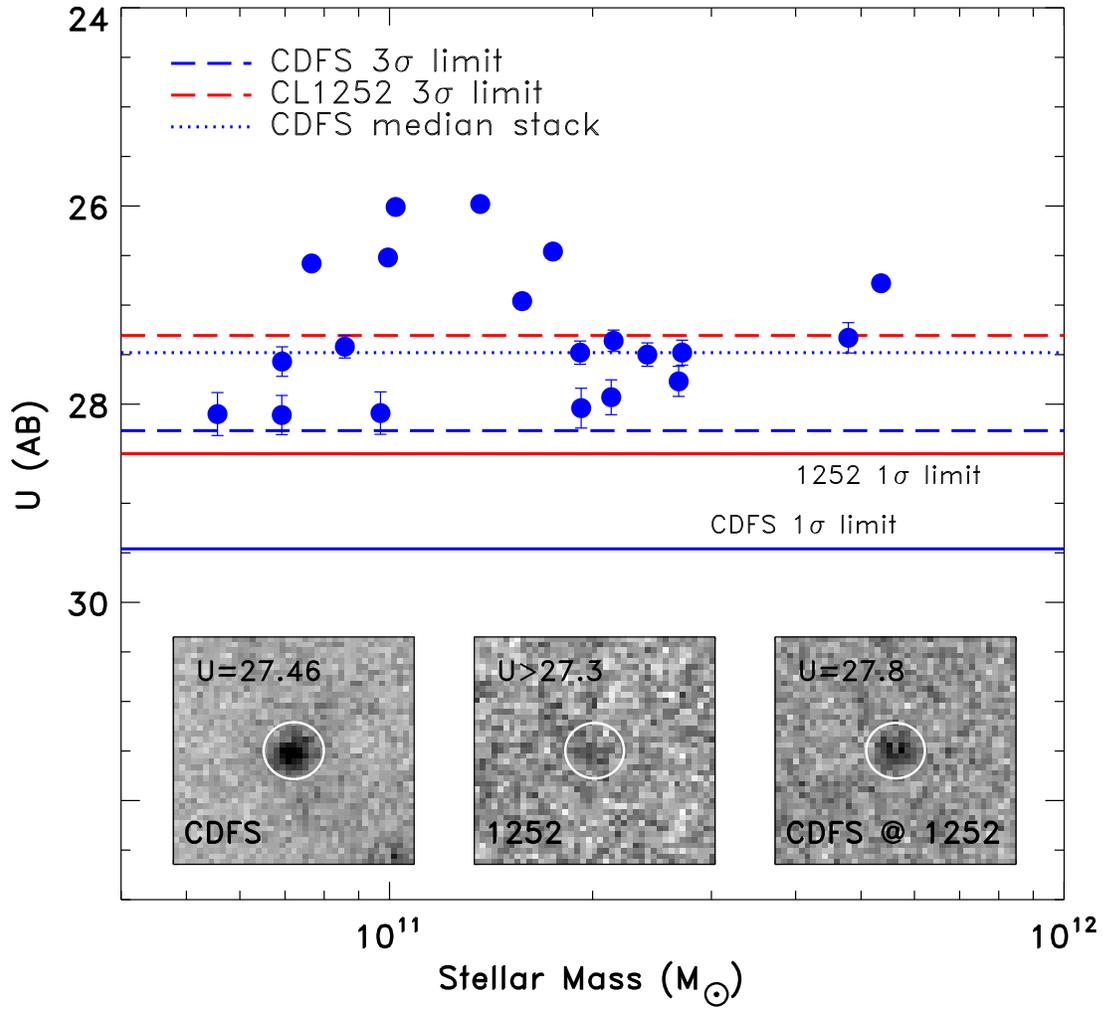}
\caption{Stellar  Mass vs.   $U$-band observed  magnitude (rest-frame
  $\sim 1700  \AA $) for the  ETG detected in the  field (filled blue
  circles).  Solid lines represent the 1$\sigma$ (1'' radius aperture)
  limiting magnitudes of our  data-sets respectively (blue for the field,
  red for  the cluster) Dashed  lines represent the  3$\sigma$ limits.
  The blue dotted line indicates  the observed magnitude of the median
  stack of  the field  ETG $U$-band images.   An image of  the median
  stack of  the field ETG  $U$-band observations is displayed  in the
  bottom-left corner.  The  median stack of the cluster  ones is shown
  in the middle.   The image at the bottom-right  corner shows instead
  the CDFS  stack, simulated  in the actual  CL1252 data. (see  text for
  more details).}
\label{uband}
\end{figure*}

\acknowledgments A.R.  is grateful  to Roderik Overzier, Arjen van der
Wel,  and  Loredana Vetere  for  useful  discussions.   A.R.  is  also
grateful   to   Andrew  Zirm   for   providing   the   data  used   in
Fig.~\ref{andrew}.

\email{aastex-help@aas.org}.

%% To help institutions obtain information on the effectiveness of their
%% telescopes, the AAS Journals has created a group of keywords for telescope
%% facilities. A common set of keywords will make these types of searches
%% significantly easier and more accurate. In addition, they will also be
%% useful in linking papers together which utilize the same telescopes
%% within the framework of the National Virtual Observatory.
%% See the AASTeX Web site at http://www.journals.uchicago.edu/AAS/AASTeX
%% for information on obtaining the facility keywords.

%% After the acknowledgments section, use the following syntax and the
%% \facility{} macro to list the keywords of facilities used in the research
%% for the paper.  Each keyword will be checked against the master list during
%% copy editing.  Individual instruments or configurations can be provided 
%% in parentheses, after the keyword, but they will not be verified.

%{\it Facilities:} \facility{Nickel}, \facility{HST (STIS)}, \facility{CXO (ASIS)}.

%% Appendix material should be preceded with a single \appendix command.
%% There should be a \section command for each appendix. Mark appendix
%% subsections with the same markup you use in the main body of the paper.

%% Each Appendix (indicated with \section) will be lettered A, B, C, etc.
%% The equation counter will reset when it encounters the \appendix
%% command and will number appendix equations (A1), (A2), etc.

\clearpage

\clearpage

\clearpage

\clearpage

\clearpage

\clearpage

\clearpage

\clearpage

\clearpage

\clearpage


\begin{thebibliography}{88}
%\expandafter\ifx\csname natexlab\endcsname\relax\def\natexlab#1{#1}\fi

\bibitem[Adelberger et al.(2005)]{Adelberger05} Adelberger, K.~L., 
Steidel, C.~C., Pettini, M., Shapley, A.~E., Reddy, N.~A., 
\& Erb, D.~K.\ 2005, \apj, 619, 697 

\bibitem[Balogh et al.(2002)]{Balogh02} Balogh, M.~L., et al.\ 
2002, \apj, 566, 123 


\bibitem[Baugh et al.(1996)]{Baugh96} Baugh, C.~M., Cole, S., 
\& Frenk, C.~S.\ 1996, \mnras, 283, 1361 

\bibitem[Bell et al.(2003)]{Bell03} Bell, E.~F., McIntosh, 
D.~H., Katz, N., \& Weinberg, M.~D.\ 2003, \apjs, 149, 289 

\bibitem[Bell et al.(2005)]{Bell05} Bell, E.~F., et al.\ 2005, 
\apj, 625, 23 

\bibitem[Bernardi et al.(1998)]{Bernardi98} Bernardi, M., Renzini, 
A., da Costa, L.~N., Wegner, G., Alonso, M.~V., Pellegrini, P.~S., 
Rit{\'e}, C., \& Willmer, C.~N.~A.\ 1998, \apjl, 508, L143 

\bibitem[Bernardi et al.(2003)]{Bernardi03} Bernardi, M., et al.\ 
2003, \aj, 125, 1866 

\bibitem[Bernardi et al.(2006)]{Bernardi06} Bernardi, M., Nichol, R.~C., Sheth, R.~K., Miller, C.~J., \& Brinkmann, J.\ 2006, \aj, 131, 1288

\bibitem[{{Blakeslee} {et~al.}(2003){Blakeslee}, {Franx}, {Postman}, {Rosati},
  {Holden}, {Illingworth}, {Ford}, {Cross}, {Gronwall}, {Ben{\'{\i}}tez},
  {Bouwens}, {Broadhurst}, {Clampin}, {Demarco}, {Golimowski}, {Hartig},
  {Infante}, {Martel}, {Miley}, {Menanteau}, {Meurer}, {Sirianni}, \&
  {White}}]{Blake03}
{Blakeslee}, J.~P., {Franx}, M., {Postman}, M., {et~al.} 2003, \apjl, 596, L143


\bibitem[Blakeslee et al.(2006)]{Blakeslee06} Blakeslee, J.~P., et 
al.\ 2006, \apj, 644, 30 

\bibitem[Bower et al.(1992)]{Bower92} Bower, R.~G., Lucey, 
J.~R., \& Ellis, R.~S.\ 1992, \mnras, 254, 601 

\bibitem[Boylan-Kolchin et al.(2006)]{Boylan06} Boylan-Kolchin, 
M., Ma, C.-P., \& Quataert, E.\ 2006, \mnras, 369, 1081 

\bibitem[Brinchmann et al.(2004)]{Brinchmann04} Brinchmann, J., 
Charlot, S., White, S.~D.~M., Tremonti, C., Kauffmann, G., Heckman, T., 
\& Brinkmann, J.\ 2004, \mnras, 351, 1151 

\bibitem[{{Bruzual} \& {Charlot}(2003)}]{BC03}
{Bruzual}, G. \& {Charlot}, S. 2003, \mnras, 344, 1000

\bibitem[{{Capaccioli}(1989)}]{Cap89}
{Capaccioli}, M. 1989, in World of Galaxies (Le Monde des Galaxies), 208--227

\bibitem[Cardelli et al.(1988)]{Cardelli88} Cardelli, J.~A., 
Clayton, G.~C., \& Mathis, J.~S.\ 1988, \apjl, 329, L33 

\bibitem[{{Cardelli} {et~al.}(1989){Cardelli}, {Clayton}, \&
  {Mathis}}]{Cardelli89}
{Cardelli}, J.~A., {Clayton}, G.~C., \& {Mathis}, J.~S. 1989, \apj, 345, 245

\bibitem[{{Cimatti} {et~al.}(2002){Cimatti}, {Pozzetti}, {Mignoli}, {Daddi},
  {Menci}, {Poli}, {Fontana}, {Renzini}, {Zamorani}, {Broadhurst}, {Cristiani},
  {D'Odorico}, {Giallongo}, \& {Gilmozzi}}]{Cimatti02}
{Cimatti}, A., {Pozzetti}, L., {Mignoli}, M., {et~al.} 2002, \aap, 391, L1

\bibitem[{{Cimatti} {et~al.}(2004){Cimatti}, {Daddi}, {Renzini}, {Cassata},
  {Vanzella}, {Pozzetti}, {Cristiani}, {Fontana}, {Rodighiero}, {Mignoli}, \&
  {Zamorani}}]{Cimatti04}
{Cimatti}, A., {Daddi}, E., {Renzini}, A., {et~al.} 2004, \nat, 430, 184

\bibitem[Ciotti \& van Albada(2001)]{Ciotti01} Ciotti, L., \& van Albada, T.~S.\ 2001, \apjl, 552, L13 

\bibitem[Clemens et al.(2006)]{Clemens06} Clemens, M.~S., 
Bressan, A., Nikolic, B., Alexander, P., Annibali, F., 
\& Rampazzo, R.\ 2006, \mnras, 370, 702 

\bibitem[Cowie et al.(1996)]{Cowie96} Cowie, L.~L., Songaila, 
A., Hu, E.~M., \& Cohen, J.~G.\ 1996, \aj, 112, 839 

\bibitem[Daddi et al.(2005)]{Daddi05} Daddi, E., et al.\ 2005, 
\apj, 626, 680 

\bibitem[De Lucia et al.(2004)]{Delucia04} De Lucia, G., 
Kauffmann, G., \& White, S.~D.~M.\ 2004, \mnras, 349, 1101 

\bibitem[De Lucia et al.(2006)]{Delucia06} De Lucia, G., 
Springel, V., White, S.~D.~M., Croton, D., \& Kauffmann, G.\ 2006, \mnras, 366, 499 

\bibitem[Demarco et al.(2007)]{Demarco07} Demarco, R., et al.\ 
2007, \apj, 663, 164 


\bibitem[Diaferio et al.(2001)]{Diaferio01} Diaferio, A., 
Kauffmann, G., Balogh, M.~L., White, S.~D.~M., Schade, D., 
\& Ellingson, E.\ 2001, \mnras, 323, 999 

\bibitem[De Propris et al.(2007)]{Depropris07} De Propris, R., 
Stanford, S.~A., Eisenhardt, P.~R., Holden, B.~P., 
\& Rosati, P.\ 2007, \aj, 133, 2209 

\bibitem[{{di Serego Alighieri} {et~al.}(2005){di Serego Alighieri}, {Vernet},
  {Cimatti}, {Lanzoni}, {Cassata}, {Ciotti}, {Daddi}, {Mignoli}, {Pignatelli},
  {Pozzetti}, {Renzini}, {Rettura}, \& {Zamorani}}]{dSA05}
{di Serego Alighieri}, S., {Vernet}, J., {Cimatti}, A., {et~al.} 2005, \aap,
  442, 125

\bibitem[di Serego Alighieri et al.(2006)]{dSA06} di Serego Alighieri,
  S., Lanzoni, B., \& J{\o}rgensen, I.\ 2006, \apjl, 652, L145

\bibitem[{{Djorgovski} \& {Davis}(1987)}]{Djorgo87}
{Djorgovski}, S. \& {Davis}, M. 1987, \apj, 313, 59

\bibitem[Dressler et al.(1997)]{Dressler97} Dressler, A., et al.\ 
1997, \apj, 490, 577 

\bibitem[{{Fioc} \& {Rocca-Volmerange}(1997)}]{Fioc97}
{Fioc}, M. \& {Rocca-Volmerange}, B. 1997, \aap, 326, 950

\bibitem[{{Fontana} {et~al.}(2004){Fontana}, {Pozzetti}, {Donnarumma},
  {Renzini}, {Cimatti}, {Zamorani}, {Menci}, {Daddi}, {Giallongo}, {Mignoli},
  {Perna}, {Salimbeni}, {Saracco}, {Broadhurst}, {Cristiani}, {D'Odorico}, \&
  {Gilmozzi}}]{Fontana04}
{Fontana}, A., {Pozzetti}, L., {Donnarumma}, I., {et~al.} 2004, \aap, 424, 23

\bibitem[Franx et al.(2003)]{Franx03} Franx, M., et al.\ 2003, 
\apjl, 587, L79 

\bibitem[{{Gavazzi} {et~al.}(1996){Gavazzi}, {Pierini}, \&
  {Boselli}}]{Gavazzi96}
{Gavazzi}, G., {Pierini}, D., \& {Boselli}, A. 1996, \aap, 312, 397

\bibitem[Gawiser et al.(2006)]{Gawiser06} Gawiser, E., et al.\ 
2006, \apjs, 162, 1 

\bibitem[{{Giavalisco} {et~al.}(2004){Giavalisco}, {Ferguson}, {Koekemoer},
  {Dickinson}, {Alexander}, {Bauer}, {Bergeron}, {Biagetti}, {Brandt},
  {Casertano}, {Cesarsky}, {Chatzichristou}, {Conselice}, {Cristiani}, {Da
  Costa}, {Dahlen}, {de Mello}, {Eisenhardt}, {Erben}, {Fall}, {Fassnacht},
  {Fosbury}, {Fruchter}, {Gardner}, {Grogin}, {Hook}, {Hornschemeier}, {Idzi},
  {Jogee}, {Kretchmer}, {Laidler}, {Lee}, {Livio}, {Lucas}, {Madau},
  {Mobasher}, {Moustakas}, {Nonino}, {Padovani}, {Papovich}, {Park},
  {Ravindranath}, {Renzini}, {Richardson}, {Riess}, {Rosati}, {Schirmer},
  {Schreier}, {Somervile}, {Spinrad}, {Stern}, {Stiavelli}, {Strolger}, {Urry},
  {Vandame}, {Williams}, \& {Wolf}}]{Giava04}
{Giavalisco}, M., {Ferguson}, H.~C., {Koekemoer}, A.~M., {et~al.} 2004, \apjl,
  600, L93

\bibitem[Gobat et al.(2008)]{Gobat08} Gobat, R., Rosati, P., Strazzullo, V., Rettura, A., Demarco, R., \& Nonino, M. \ 2008, \aap, submitted 

\bibitem[Graham 
\& Guzm{\'a}n(2003)]{2003AJ....125.2936G} Graham, A.~W., \& Guzm{\'a}n, R.\ 2003, \aj, 125, 2936 

\bibitem[Gunn \& Gott(1972)]{Gunn72} Gunn, J.~E., \& Gott, J.~R.~I.\ 1972, \apj, 176, 1 

\bibitem[Jorgensen et al.(1996)]{Jorgensen96} Jorgensen, I., Franx, 
M., \& Kjaergaard, P.\ 1996, \mnras, 280, 167 

\bibitem[J{\o}rgensen et al.(2006)]{Jorgensen06} J{\o}rgensen, I., 
Chiboucas, K., Flint, K., Bergmann, M., Barr, J., 
\& Davies, R.\ 2006, \apjl, 639, L9 

\bibitem[Juneau et al.(2005)]{Juneau05} Juneau, S., et al.\ 
2005, \apjl, 619, L135

\bibitem[Hilton et al.(2007)]{Hilton07} Hilton, M., et al.\ 
2007, \apj, 670, 1000 

\bibitem[{{Holden} {et~al.}(2005){Holden}, {van der Wel}, {Franx},
  {Illingworth}, {Blakeslee}, {van Dokkum}, {Ford}, {Magee}, {Postman}, {Rix},
  \& {Rosati}}]{Hol05}
{Holden}, B.~P., {van der Wel}, A., {Franx}, M., {et~al.} 2005, \apjl, 620, L83

\bibitem[Kauffmann \& Charlot(1998)]{Kauffmann98} Kauffmann, G., \& Charlot, S.\ 1998, \mnras, 294, 705 

\bibitem[{{Kauffmann} {et~al.}(2003){Kauffmann}, {Heckman}, {White}, {Charlot},
  {Tremonti}, {Peng}, {Seibert}, {Brinkmann}, {Nichol}, {SubbaRao}, \&
  {York}}]{Kauffmann03}
{Kauffmann}, G., {Heckman}, T.~M., {White}, S.~D.~M., {et~al.} 2003, \mnras,
  341, 54

\bibitem[Kauffmann et al.(2007)]{Kauffmann07} Kauffmann, G., et 
al.\ 2007, \apjs, 173, 357 

\bibitem[Kaviraj et al.(2005)]{Kaviraj05} Kaviraj, S., Devriendt, 
J.~E.~G., Ferreras, I., \& Yi, S.~K.\ 2005, \mnras, 360, 60 

\bibitem[Kaviraj et al.(2007)]{Kaviraj07} Kaviraj, S., et al.\ 
2007, ArXiv e-prints, 709, arXiv:0709.0806 

\bibitem[Kochanek et al.(2000)]{Kochanek00} Kochanek, C.~S., et 
al.\ 2000, \apj, 543, 131 

\bibitem[Khochfar \& Burkert(2003)]{Khochfar03} Khochfar, S., \& Burkert, A.\ 2003, \apjl, 597, L117 

\bibitem[Khochfar 
\& Silk(2006)]{Khochfar06} Khochfar, S., \& Silk, J.\ 2006, \apjl, 648, L21 

\bibitem[Kodama 
\& Bower(2001)]{Kodama01} Kodama, T., \& Bower, R.~G.\ 2001, \mnras, 321, 18 


\bibitem[Kodama et al.(2007)]{Kodama07} Kodama, T., Tanaka, I., 
Kajisawa, M., Kurk, J., Venemans, B., De Breuck, C., Vernet, J., 
\& Lidman, C.\ 2007, \mnras, 377, 1717 


\bibitem[Kormendy(1977)]{kor77} Kormendy, J.\ 1977, \apj, 218, 333 

\bibitem[Kriek et al.(2006)]{Kriek06} Kriek, M., et al.\ 2006, 
\apj, 645, 44 

\bibitem[Krist(1995)]{Krist95} Krist, J.\ 1995, Astronomical 
Data Analysis Software and Systems IV, 77, 349 

\bibitem[Kurk et al.(2004)]{Kurk04} Kurk, J.~D., Pentericci, L., R{\"o}ttgering, H.~J.~A., \& Miley, G.~K.\ 2004, \aap, 428, 793 

\bibitem[La Barbera et al.(2003)]{LaBarbera03} La Barbera, F., 
Busarello, G., Merluzzi, P., Massarotti, M., 
\& Capaccioli, M.\ 2003, \apj, 595, 127 


\bibitem[Le Borgne et al.(2006)]{Leborgne06} Le Borgne, D., et 
al.\ 2006, \apj, 642, 48

\bibitem[{{Lidman} {et~al.}(2004){Lidman}, {Rosati}, {Demarco}, {Nonino},
  {Mainieri}, {Stanford}, \& {Toft}}]{Lidman04}
{Lidman}, C., {Rosati}, P., {Demarco}, R., {et~al.} 2004, \aap, 416, 829


\bibitem[Labb{\'e} et al.(2005)]{Labbe05} Labb{\'e}, I., et 
al.\ 2005, \apjl, 624, L81 

\bibitem[Longhetti et al.(2007)]{Longhetti07} Longhetti, M., et 
al.\ 2007, \mnras, 374, 614 

\bibitem[{{Maraston}(1998)}]{Maraston98}
{Maraston}, C. 1998, \mnras, 300, 872

\bibitem[{{Maraston}(2005)}]{Maraston05}
{Maraston}, C. 2005, \mnras, 362, 799

\bibitem[{{Marleau} \& {Simard}(1998)}]{Marleau98}
{Marleau}, F.~R. \& {Simard}, L. 1998, \apj, 507, 585


\bibitem[McIntosh et al.(2005)]{McIntosh05} McIntosh, D.~H., et 
al.\ 2005, \apj, 632, 191 


\bibitem[Mei et al.(2006a)]{Mei06a} Mei, S., et al.\ 2006, 
\apj, 644, 759 

\bibitem[Mei et al.(2006b)]{Mei06b} Mei, S., et al.\ 2006, 
\apj, 639, 81 

\bibitem[Menci et al.(2008)]{Menci08} Menci, N., Rosati, P., Gobat, R., Strazzullo, V., Rettura, Mei, S., \& Demarco, R. \ 2008, \apj, submitted 

\bibitem[Moore et al.(1996)]{Moore96} Moore, B., Katz, N., 
Lake, G., Dressler, A., \& Oemler, A.\ 1996, \nat, 379, 613 

\bibitem[Moore et al.(1998)]{Moore98} Moore, B., Lake, G., 
\& Katz, N.\ 1998, \apj, 495, 139 

\bibitem[Moore et al.(1999)]{Moore99} Moore, B., Ghigna, S., 
Governato, F., Lake, G., Quinn, T., Stadel, J., 
\& Tozzi, P.\ 1999, \apjl, 524, L19 

\bibitem[Nipoti et al.(2003)]{Nipoti03} Nipoti, C., Londrillo, 
P., \& Ciotti, L.\ 2003, \mnras, 342, 501 

\bibitem[Nonino  et al.(2008)]{Nonino08}  Nonino, M.,  Rosati, Rettura,
  A. et al.\ 2008, in preparation

\bibitem[Papovich et al.(2006)]{Papovich06} Papovich, C., et al.\ 
2006, \apj, 640, 92 

\bibitem[Renzini(2006)]{Renzini06} Renzini, A.\ 2006, \araa, 44, 141 

\bibitem[Rettura et al.(2006)]{Rettura06} Rettura, A., et al.\ 
2006, \aap, 458, 717 

\bibitem[{{Rosati} {et~al.}(2004){Rosati}, {Tozzi}, {Ettori}, {Mainieri},
  {Demarco}, {Stanford}, {Lidman}, {Nonino}, {Borgani}, {Della Ceca},
  {Eisenhardt}, {Holden}, \& {Norman}}]{Rosati04}
{Rosati}, P., {Tozzi}, P., {Ettori}, S., {et~al.} 2004, \aj, 127, 230

\bibitem[{{Salpeter}(1955)}]{Salpeter55}
{Salpeter}, E.~E. 1955, \apj, 121, 161

\bibitem[S{\'a}nchez-Bl{\'a}zquez et 
al.(2006)]{Sanchez06} S{\'a}nchez-Bl{\'a}zquez, P., Gorgas, J., Cardiel, N., \& Gonz{\'a}lez, J.~J.\ 2006, \aap, 457, 809 

\bibitem[{{Saracco} {et~al.}(2004){Saracco}, {Longhetti}, {Giallongo},
  {Arnouts}, {Cristiani}, {D'Odorico}, {Fontana}, {Nonino}, \&
  {Vanzella}}]{Saracco04}
{Saracco}, P., {Longhetti}, M., {Giallongo}, E., {et~al.} 2004, \aap, 420, 125

\bibitem[Sawicki   \&  Thompson(2006)]{Sawicki06}   Sawicki,   M.,  \&
  Thompson, D.\ 2006, \apj, 642, 653


\bibitem[{{S\'ersic}(1968)}]{Sersic68}
{Sersic}, J.~L. 1968, {Atlas de galaxias australes} (Cordoba, Argentina: Obs.
  Astronomico, 1968)

\bibitem[Shen et al.(2003)]{Shen03} Shen, S., Mo, H.~J., 
White, S.~D.~M., Blanton, M.~R., Kauffmann, G., Voges, W., Brinkmann, J., 
\& Csabai, I.\ 2003, \mnras, 343, 978 

\bibitem[{{Simard}(1998)}]{Simard98}
{Simard}, L. 1998, in ASP Conf. Ser. 145, 108--+

\bibitem[{{Somerville} {et~al.}(2004){Somerville}, {Moustakas}, {Mobasher},
  {Gardner}, {Cimatti}, {Conselice}, {Daddi}, {Dahlen}, {Dickinson},
  {Eisenhardt}, {Lotz}, {Papovich}, {Renzini}, \& {Stern}}]{Somerville04}
{Somerville}, R.~S., {Moustakas}, L.~A., {Mobasher}, B., {et~al.} 2004, \apjl,
  600, L135

\bibitem[{{Spergel} {et~al.}(2003){Spergel}, {Verde}, {Peiris}, {Komatsu},
  {Nolta}, {Bennett}, {Halpern}, {Hinshaw}, {Jarosik}, {Kogut}, {Limon},
  {Meyer}, {Page}, {Tucker}, {Weiland}, {Wollack}, \& {Wright}}]{Spergel03}
{Spergel}, D.~N., {Verde}, L., {Peiris}, H.~V., {et~al.} 2003, \apjs, 148, 175

\bibitem[Strazzullo et al.(2006)]{Strazzullo06} Strazzullo, V., et 
al.\ 2006, \aap, 450, 909 


\bibitem[Steidel et al.(2005)]{Steidel05} Steidel, C.~C., 
Adelberger, K.~L., Shapley, A.~E., Erb, D.~K., Reddy, N.~A., 
\& Pettini, M.\ 2005, \apj, 626, 44 

\bibitem[Tanaka et al.(2005)]{Tanaka05} Tanaka, M., Kodama, T., 
Arimoto, N., Okamura, S., Umetsu, K., Shimasaku, K., Tanaka, I., 
\& Yamada, T.\ 2005, \mnras, 362, 268 

\bibitem[Thomas et al.(2004)]{Thomas04} Thomas, D., Maraston, 
C., \& Korn, A.\ 2004, \mnras, 351, L19 

\bibitem[{{Thomas} {et~al.}(2005){Thomas}, {Saglia}, {Bender}, {Thomas},
  {Gebhardt}, {Magorrian}, {Corsini}, \& {Wegner}}]{Thomas05}
{Thomas}, J., {Saglia}, R.~P., {Bender}, R., {et~al.} 2005, \mnras, 360, 1355


\bibitem[Toft et al.(2004)]{Toft04} Toft, S., Mainieri, V., Rosati, P., Lidman, C., Demarco, R., Nonino, M., \& Stanford, S.~A.\ 2004, \aap, 422, 29 


\bibitem[Toft et al.(2007)]{Toft07} Toft, S., et al.\ 2007, 
\apj, 671, 285 

\bibitem[Trager et al.(2000)]{Trager00} Trager, S.~C., Faber, 
S.~M., Worthey, G., \& Gonz{\'a}lez, J.~J.\ 2000, \aj, 119, 1645 

\bibitem[Treu et al.(1999)]{Treu99} Treu, T., Stiavelli, M., 
Casertano, S., M{\o}ller, P., \& Bertin, G.\ 1999, \mnras, 308, 1037 

\bibitem[Treu et al.(2001)]{Treu01} Treu, T., Stiavelli, M., 
Bertin, G., Casertano, S., \& M{\o}ller, P.\ 2001, \mnras, 326, 237 

\bibitem[{{Treu} {et~al.}(2005){Treu}, {Ellis}, {Liao}, {van Dokkum}, {Tozzi},
  {Coil}, {Newman}, {Cooper}, \& {Davis}}]{Treu05}
{Treu}, T., {Ellis}, R.~S., {Liao}, T.~X., {et~al.} 2005, \apj, 633, 174

\bibitem[Trujillo et al.(2004)]{Trujillo04} Trujillo, I., et al.\ 
2004, \apj, 604, 521 

\bibitem[Trujillo et al.(2006)]{Trujillo06} Trujillo, I., et al.\ 
2006, \apj, 650, 18 

\bibitem[Trujillo et al.(2007)]{Trujillo07} Trujillo, I., 
Conselice, C.~J., Bundy, K., Cooper, M.~C., Eisenhardt, P., 
\& Ellis, R.~S.\ 2007, \mnras, 382, 109 

\bibitem[van der Wel et al.(2005)]{vdW05} van der Wel, A., 
Franx, M., van Dokkum, P.~G., Rix, H.-W., Illingworth, G.~D., 
\& Rosati, P.\ 2005, \apj, 631, 145 

\bibitem[van der Wel et al.(2006)]{vdW06} van der Wel, A., 
Franx, M., Wuyts, S., van Dokkum, P.~G., Huang, J., Rix, H.-W., 
\& Illingworth, G.~D.\ 2006, \apj, 652, 97 

\bibitem[{{van Dokkum} \& {Franx}(1996)}]{vandokkum96}
{van Dokkum}, P.~G. \& {Franx}, M. 1996, \mnras, 281, 985

\bibitem[{{van Dokkum} {et~al.}(1998){van Dokkum}, {Franx}, {Kelson}, \&
  {Illingworth}}]{vandokkum98}
{van Dokkum}, P.~G., {Franx}, M., {Kelson}, D.~D., \& {Illingworth}, G.~D.
  1998, \apjl, 504, L17+

\bibitem[van Dokkum 
\& Franx(2001)]{vandokkum01} van Dokkum, P.~G., \& Franx, M.\ 2001, \apj, 553, 90 
\bibitem[van Dokkum \& Stanford(2003)]{vandokkumstanford03} van Dokkum, P.~G., \& Stanford, S.~A.\ 2003, \apj, 585, 78 

\bibitem[van Dokkum \& van der Marel(2007)]{vandokkum07} van Dokkum, P.~G., \& van der Marel, R.~P.\ 2007, \apj, 655, 30 

\bibitem[Vanzella et al.(2005)]{Vanzella05} Vanzella, E., et al.\ 
2005, \aap, 434, 53 

\bibitem[Vanzella et al.(2006)]{Vanzella06} Vanzella, E., et al.\ 
2006, \aap, 454, 423 

\bibitem[Wolf et 
al.(2004)]{Wolf04} Wolf, C., et al.\ 2004, \aap, 421, 913 

\bibitem[Worthey(1994)]{Worthey94} Worthey, G.\ 1994, \apjs, 95, 
107 


\bibitem[Yi et al.(1999)]{Yi99} Yi, S., Lee, Y.-W., Woo, 
J.-H., Park, J.-H., Demarque, P., \& Oemler, A.~J.\ 1999, \apj, 513, 128 

\bibitem[Yi et al.(1997)]{Yi97} Yi, S., Demarque, P.,  \& Oemler, A.~J.\ 1997, \apj, 486, 201 


\bibitem[Zirm et al.(2007)]{Zirm07} Zirm, A.~W., et al.\ 2007, 
\apj, 656, 66 

\bibitem[Zirm et al.(2008)]{Zirm08} Zirm, A.~W., et al.\ 2008, 
ArXiv e-prints, 802, arXiv:0802.2095 

\end{thebibliography}
\end{document}